\documentclass[11pt,a4paper]{article}

\usepackage{epsfig}
\usepackage{amsmath}
\usepackage{amssymb}
\usepackage{graphicx}
\usepackage{float}
\usepackage{subfig}
\usepackage{vmargin}
\usepackage{mathrsfs}
\usepackage{mathbbol}
\usepackage{lscape}
\usepackage{color}
\usepackage[citecolor=blue,colorlinks=true,linkcolor=blue]{hyperref}
\usepackage[round]{natbib}
\usepackage[parfill]{parskip}
\setmargrb{20mm}{20mm}{20mm}{20mm}
\usepackage[noblocks]{authblk}

\usepackage{tikz}
\usetikzlibrary{matrix}
\usepackage{pgfplots}

\newcommand{\be}{\begin{equation}}
\newcommand{\ee}{\end{equation}}
\newcommand{\ba}{\begin{equation} \begin{aligned}}
\newcommand{\ea}{\end{aligned} \end{equation}}

\newcommand{\dint}[1]{\mathrm{d}#1}

\title{\sc Inferring Risks of Coronavirus Transmission from Community Household
Data}
\author[1,2,3,*]{Thomas House}
\author[1]{Heather Riley}
\author[1,3]{Lorenzo Pellis}
\author[4,5,6]{Koen B.~Pouwels}
\author[7]{Sebastian Bacon}
\author[8]{Arturas Eidukas}
\author[8]{Kaveh Jahanshahi}
\author[9]{Rosalind M.~Eggo}
\author[4,5,10,11]{A.~Sarah Walker}
\affil[1]{Department of Mathematics, University of Manchester, Manchester, UK}
\affil[2]{IBM Research, Hartree Centre, Daresbury, UK}
\affil[3]{The Alan Turing Institute for Data Science and Artificial Intelligence, London, UK}
\affil[4]{Nuffield Department of Medicine, University of Oxford, Oxford, UK}
\affil[5]{The National Institute for Health Research Health Protection Research
Unit in Healthcare Associated Infections and Antimicrobial Resistance at the
University of Oxford, Oxford, UK}
\affil[6]{Health Economics Research Centre, Nuffield Department of Population
Health, University of Oxford, Oxford, UK}
\affil[7]{The DataLab, Nuffield Department of Primary Care Health Sciences,
University of Oxford, Oxford, UK}
\affil[8]{Data Science Campus, Office for National Statistics (ONS)}
\affil[9]{Centre for Mathematical Modelling of Infectious Diseases, London
School of Hygiene and Tropical Medicine, London, UK}
\affil[10]{The National Institute for Health Research Oxford Biomedical Research
Centre, University of Oxford, Oxford, UK}
\affil[11]{MRC Clinical Trials Unit at UCL, UCL, London, UK}
\affil[*]{Corresponding Author: thomas.house@manchester.ac.uk}

\date{}

\begin{document}


\maketitle

\begin{abstract}
\noindent{}The response of many governments to the COVID-19 pandemic has
involved measures to control within- and between-household transmission,
providing motivation to improve understanding of the absolute and relative
risks in these contexts. Here, we perform exploratory, residual-based, and
transmission-dynamic household analysis of the Office for National Statistics
(ONS) COVID-19 Infection Survey (CIS) data from 26 April 2020 to 15 July 2021
in England. This provides evidence for: (i) temporally varying rates of
introduction of infection into households broadly following the trajectory of
the overall epidemic and vaccination programme; (ii) Susceptible-Infectious
Transmission Probabilities (SITPs) of within-household transmission in the
15-35\% range; (iii) the emergence of the Alpha and Delta variants, with the
former being around 50\% more infectious than wildtype and 35\% less infectious
than Delta within households; (iv) significantly (in the range 25-300\%) more
risk of bringing infection into the household for workers in patient-facing
roles pre-vaccine; (v) increased risk for secondary school-age children of
bringing the infection into the household when schools are open; (vi) increased
risk for primary school-age children of bringing the infection into the
household when schools were open since the emergence of new variants.
\end{abstract}

\section{Introduction}

\subsection{Analysis of household infection data}

Households have often played an important role in infectious disease
epidemiology, with policies in place and under consideration in the UK to
reduce both within- and between-household transmission \citep{SAGE:2021}. This
is because the close, repeated nature of contact within the household means
that within-household transmission of infectious disease is common. Also, most
of the population lives in relatively small, stable households \citep{FH:2019}.
From the point of view of scientific research, the household is a natural unit
for epidemiological data collection and households are small enough to allow
for explicit solution of relatively complex transmission models. Some of the
earliest work in this field was carried out by Reed and Frost, whose model was
first described in the literature by \citet{Abbey:1952} in a paper that
analysed transmission in boarding schools and other closed populations. Frost's
1928 lecture was later published posthumously \citep{Frost:1976}, with a
re-analysis of his original household dataset from the 1918 influenza pandemic
carried out using modern computational and modelling approaches by
\citet{Fraser:2011}.

Subsequent important contributions were made in empirical studies of
transmission in households, for example the highly influential study of
childhood diseases by \citet{HopeSimpson:1952}, and epidemic theory based on
the analyses of discrete- and continuous-time Markovian epidemics presented by
\citet{Bailey:1957}. A key development was the solution by \citet{Ball:1986} of
the final size distribution of a random epidemic in a household without
requiring Markovian recovery from infection, which then enabled statistical
analyses of household infection data such as that by \citet{Addy:1991}. Still
further progress is possible due to the use of modern computational methods,
particularly Monte Carlo approaches, to augment datasets
\citep{ONeill:1999,Cauchemez:2004,Demiris:2005} or to avoid likelihood
calculations \citep{Neal:2012}.

Continued methodological developments and data availability have enabled
increasingly sophisticated inferences to be drawn from household studies of
respiratory pathogens, dealing with for example interactions between adults and
children \citep{vanBoven:2010}, case ascertainment \citep{House:2012},
interactions between strains \citep{Kombe:2019}, and details of family
structure \citep{Endo:2019}.  During the current pandemic, there have been
numerous household studies \citep{Madewell:2020}, with three recently published
studies being notable for combining fitting of a transmission model with
significant differentiation of risks being those of \citet{Dattner:2021},
\citet{Li:2021} and \citet{Reukers:2021}.

\subsection{Context for this study}

The severe acute respiratory syndrome coronavirus 2 (SARS-CoV-2) emerged in the
human population in late 2019 and the WHO declared a pandemic in March 2020
\citep{WHO:2020}. Early in the pandemic, it became clear that risks of
transmission, mortality and morbidity from the associated coronavirus disease
(COVID-19) were highly heterogeneous with age \citep{Davies:2020}, and also
that work in patient-facing roles was associated with increased risk of
positivity in the community \citep{Pouwels:2021} as would be expected given the
risks of healthcare-associated transmission \citep{Bhattacharya:2021}.

During the period of the study, there have been two major `sweeps' in the UK,
during which a SARS-CoV-2 variant of concern (VOC) emerged and became dominant.

The first of these was PANGO lineage B.1.1.7~\citep{Rambaut:2020}, or `Alpha'
under WHO nomenclature~\citep{WHO_variants}. The first samples of this variant
were found in September 2020~\citep{Rambaut:2020}, and it was designated a VOC
on 18 December 2020~\citep{VoC_tech_briefing_1}. There is evidence for both
increased transmissibility of this variant, and increased mortality amongst
infected cases \citep{Davies:2021,Challen:2021,Grint:2021}, although
conditional on hospitalisation outcomes may not be worse~\citep{Frampton:2021}.
The second VOC to emerge was PANGO lineage B.1.617.2~\citep{Rambaut:2020pango},
or `Delta' under WHO nomenclature~\citep{WHO_variants}, which was designated a
VOC on 6 May 2021 and is at time of writing the dominant variant in the
UK~\citep{VOC_tech_briefing_10}. 

Both of these variants were relatively easy to track through the S gene target
in commonly-used polymerase chain reaction (PCR) tests, with more details on
this approach provided in \S{}\ref{sec:dod} below.

Throughout 2021, the UK rolled out a comprehensive vaccination programme with
priority given to healthcare workers, the clinically vulnerable, and then with
prioritisation by age, from oldest to youngest \citep{ONS_Vacc,PHE_Vacc}. We
will not include vaccination here at the individual level, but rather note its
overall effect on infection and transmission at different times.

Here, we apply a combination of methods, including a regression that explicitly
accounts for transmission, to the Office for National Statistics (ONS) COVID-19
Infection Survey (CIS) data from 26 April 2020 to 15 July 2021
\citep{Pouwels:2021}. We particularly consider the absolute magnitude of
transmission within and between households, as well as the associations between
these and household size, age, infection with VOCs (inferred via S gene target)
and work in patient-facing roles.

\section{Methods}

\subsection{Description of data}

\label{sec:dod}

ONS
CIS\footnote{https://www.ndm.ox.ac.uk/covid-19/covid-19-infection-survey/protocol-and-information-sheets;
ISRCTN number ISRCTN21086382; The study received ethical approval from the
South Central Berkshire B Research Ethics Committee (20/SC/0195).} has a design
based on variable levels of recruitment by region and time as required by
policy, but otherwise uniformly random selection of households from address
lists and previous ONS studies on an ongoing basis.  If verbal agreement to
participate is obtained, a study worker visits each household to take written
informed consent, which is obtained from parents/carers for those aged 2-15
years. Participants aged 10-15 years provide written assent and those under 2
years old are not eligible.

\begin{figure}[ht!]
\begin{center}
\begin{tikzpicture}
\draw[step=1.5cm,gray,very thin] (0,0) grid (12.6,7.4);
  \draw [->] (0,0) -- (12.7cm,0) node[right] {Time (weeks)} ;
  \draw [->,thick,red] (1cm,8cm) -- (1cm,6cm) ;
  \draw [-|,thick,red] (1cm,6cm) -- (4cm,6cm) ;
  \draw [->,thick,red] (2cm,6cm) -- (2cm,4.5cm) ;
  \draw [-|,thick,red] (2cm,4.5cm) -- (6.2cm,4.5cm) ;
  \draw [->,thick,red] (2.9cm,4.5cm) -- (2.9cm,3cm) ;
  \draw [-|,thick,red] (2.9cm,3cm) -- (5.7cm,3cm) ;
  \node at (0cm,-0.2cm) {$0$};
  \node at (1.5cm,-0.2cm) {$1$};
  \node at (3cm,-0.2cm) {$2$};
  \node at (4.5cm,-0.2cm) {$3$};
  \node at (6cm,-0.2cm) {$4$};
  \node at (7.5cm,-0.2cm) {$5$};
  \node at (9cm,-0.2cm) {$6$};
  \node at (10.5cm,-0.2cm) {$7$};
  \node at (12cm,-0.2cm) {$8$};
  \node[circle,fill=blue,inner sep=0pt] at (0.5cm,1.5cm) {$\times$};
  \node[circle,fill=blue,inner sep=0pt] at (0.5cm,3cm) {$\times$};
  \node[circle,fill=blue,inner sep=0pt] at (0.5cm,4.5cm) {$\times$};
  \node[circle,fill=blue,inner sep=0pt] at (0.5cm,6cm) {$\times$};
  \node[circle,fill=blue,inner sep=0pt] at (1.7cm,1.5cm) {$\times$};
  \node[circle,fill=blue,inner sep=0pt] at (1.7cm,3cm) {$\times$};
  \node[circle,fill=blue,inner sep=0pt] at (1.7cm,4.5cm) {$\times$};
  \node[circle,fill=red,inner sep=0pt] at (1.7cm,6cm) {$+$};
  \node[circle,fill=blue,inner sep=0pt] at (3.5cm,1.5cm) {$\times$};
  \node[circle,fill=red,inner sep=0pt] at (3.5cm,3cm) {$+$};
  \node[circle,fill=red,inner sep=0pt] at (3.5cm,4.5cm) {$+$};
  \node[circle,fill=red,inner sep=0pt] at (3.5cm,6cm) {$+$};
  \node[circle,fill=blue,inner sep=0pt] at (4.6cm,1.5cm) {$\times$};
  \node[circle,fill=red,inner sep=0pt] at (4.6cm,3cm) {$+$};
  \node[circle,fill=red,inner sep=0pt] at (4.6cm,4.5cm) {$+$};
  \node[circle,fill=blue,inner sep=0pt] at (4.6cm,6cm) {$\times$};
  \node[circle,fill=blue,inner sep=0pt] at (11.5cm,1.5cm) {$\times$};
  \node[circle,fill=blue,inner sep=0pt] at (11.5cm,3cm) {$\times$};
  \node[circle,fill=blue,inner sep=0pt] at (11.5cm,4.5cm) {$\times$};
  \node[circle,fill=blue,inner sep=0pt] at (11.5cm,6cm) {$\times$};
  \node at (-1cm,1.5cm) {PID$1$};
  \node at (-1cm,3cm) {PID$2$};
  \node at (-1cm,4.5cm) {PID$3$};
  \node at (-1cm,6cm) {PID$4$};
\end{tikzpicture}
\end{center}
\caption{Schematic of a hypothetical but realistic data pattern for a
four-person household in the first two months after recruitment. Each negative
test is shown as a blue circle containing $\times$ and each positive test is
shown as a red circle containing $+$. One potential route for infection coming
into and transmitting within the household is shown as through a series of red
arrows. This is not directly observed in the study design, and in fact other
transmission trees (for example one in which PID2 is infected before PID3) are
consistent with the data that would be obtained from this household.}
\label{fig:schematic}
\end{figure}
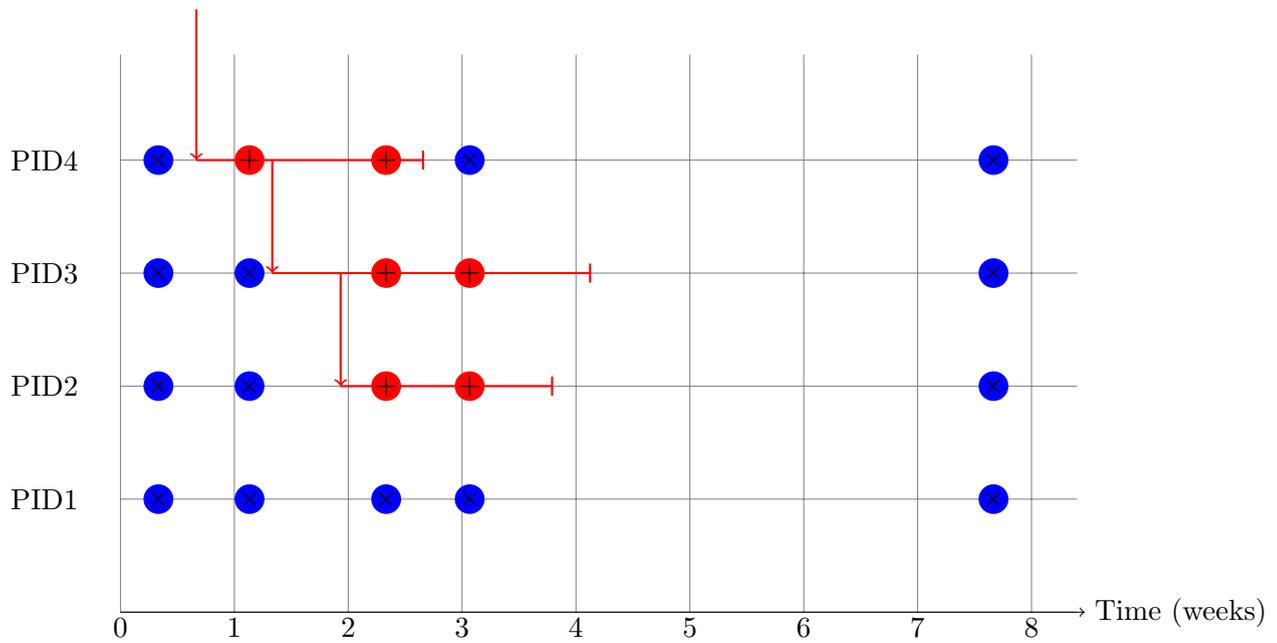

Participants are asked questions on issues including work and
age\footnote{https://www.ndm.ox.ac.uk/covid-19/covid-19-infection-survey/case-record-forms}
as well as being tested for SARS-CoV-2 infection via reverse transcription PCR
(RT-PCR).  To reduce transmission risks, participants aged 12 years and over
self-collect nose and throat swabs following study worker instructions, and
parents/carers take swabs from children aged under 12 years.  At the first
visit, participants are asked for optional consent for follow-up visits every
week for the next month, then monthly for 12 months from enrolment. The first
few weeks of a hypothetical household participating in this study are shown
schematically in Figure~\ref{fig:schematic}.

Swabs were analysed at the UK’s national Lighthouse Laboratories at Milton
Keynes and Glasgow using identical methodology. RT-PCR for three SARS-CoV-2
genes (N protein, S protein and ORF1ab) used the Thermo Fisher TaqPath RT-PCR
COVID-19 Kit, and analysed using UgenTec FastFinder 3.300.5, with an
assay-specific algorithm and decision mechanism that allows conversion of
amplification assay raw data from the ABI 7500 Fast into test results with
minimal manual intervention. Samples are called positive if at least a single
N-gene and/or ORF1ab are detected. Although S gene cycle threshold (Ct) values
are determined, S gene detection alone is not considered sufficient to call a
sample positive. 

This analysis includes all SARS-CoV-2 RT-PCR tests of nose and throat swabs
from 26 April 2020 to 15 July 2021 for English households in the ONS CIS.
We restrict our analysis to households of size 6 and under, partly for
computational reasons that we will discuss below, and partly because this
captures the overwhelming majority of households, with larger households being
atypical in various ways \citep{FH:2019}. Over 94\% of households have all
members participating, and for the remainder we treat the household as composed
of participants only.  In contrast to other studies, the households we select
constitute an approximately representative sample from the population when
stratified by date and region. The restriction to England was chosen because we
split the data into four time periods, corresponding to changing situations
about policies that are devolved (i.e.\ policies are different in Scotland,
Wales and Northern Ireland).  These time periods split the data into the
following tranches, with associated time periods and notable events (described
broadly).

\begin{itemize}
\item \textbf{Tranche 1:} 26 April 2020 to 31 August 2020; low prevalence; schools
closed; Alpha and Delta variants not emerged yet; no vaccine available.
\item \textbf{Tranche 2:} 1 September 2020 to 14 November 2020; high
prevalence; schools open; negligible Alpha variant; Delta variant not emerged yet;
no vaccine available.
\item \textbf{Tranche 3:} 15 November 2020 to 31 December 2020; high prevalence;
schools open; Alpha variant becomes dominant; Delta variant not emerged yet;
negligible vaccine coverage.
\item \textbf{Tranche 4:} 1 January 2021 to 14 February 2021; high prevalence;
schools closed (except for pre-school); Alpha variant dominant; Delta variant
not emerged yet; over 10 million first vaccine doses by end of time period.
\item \textbf{Tranche 5:} 15 February 2021 to 29 April 2021; low prevalence;
schools open; Delta variant negligible; over 35 million first and 15 million
second vaccine doses by end of time period.
\item \textbf{Tranche 6:} 30 April 2021 to 15 July 2021; high prevalence;
schools open; Delta variant becomes dominant; over 45 million first and 35
million second doses distributed by end of time period.
\end{itemize}

These properties are summarised again in Table \ref{tab:tranches}.  The
properties of the data allocated to these tranches are shown in
Table~\ref{tab:features}. Note that, while we do not include new primary
infections in households after 15 July 2021, but do include later secondary
infections in households where the primary infection happened before 15 July
2021. This is done to reduce problems with censoring.

\subsection{Mathematical representation of data}

Suppose we have a set of $n$ individuals (participants), indexed $i, j, \ldots
\in [n]$, where we use the notation $[k]$ to stand for the set of integers from
$1$ to $k$ inclusive. These individuals are members of $m$ households, and we
represent the $a$-th household using a set of individual indices $H_a$.  These
are specified such that each individual is in exactly one household, so
formally,
$$
H_a \subseteq [n], \forall a\in [m] \text{ ,} \quad
H_a \cap H_b = \varnothing, \forall a\in [m], b\in[m]\setminus\{a\} \text{ ,} \quad
\bigcup_{a=1}^{m} H_a = [n] \text{ .}
$$
The size of the $a$-th household is then $n_a = |H_a|$. The $a$-th household is
visited at a set of times $\mathcal{T}_a$, and for each $t\in \mathcal{T}_a$ we
let $\mathbf{x}_{i,t}$ be the length-$p$ feature vector (also called
covariates) associated with the $i$-th individual at time $t$, and $y_{i,t}$ be
the test result so that $y_{i,t}=1$ if the swab is positive and $y_{i,t}=0$ if
not.  Note that not all $i\in H_a$ will register a valid observation for
features and swab results for each $t\in \mathcal{T}_a$.

We let a tranche be defined by a time interval $T=[t_1, t_2)$, and the
household $H_a$ will appear in the analysis associated with the tranche $T$ if
$\mathcal{T}_a \cap T \neq \varnothing$. For the analysis that we will perform,
we require a method for associating a unique positivity and feature vector with
each individual for the duration of the tranche. Under our modelling
assumptions, the following definition of tranche positivity is most natural.
For each household $H_a$ associated with tranche $T$,
\be
\forall i \in H_a \text{ ,}\quad y_i = \begin{cases}
1 & \text{ if } \exists t, y_{i,t} = 1\ \&\
\min\{\tau|\exists j\in H_a, y_{j,\tau} = 1\} \in T \text{ ,}\\
0 & \text{ otherwise.}
\end{cases}
\ee
This means that we associate every positive in the household with the tranche
in which the first positive appears in that household. Such an approach would
need revision for a situation where individuals were infected a large number of
times (i.e.\ common reinfection) or if incidence were so high that a
significant number of households would be expected to have multiple
introductions, but we do not see these scenarios in our data. For features, the
appropriate rule will depend on the feature. An example such rule for the case
where there is only one feature $x_{i,t} \in \{0, 1\}, \forall i, t$ would be
$$
x_i = \max\{x_{i,t}| t\in \mathcal{T}_a \cap T\} \text{ ,}
$$
i.e.\ we take this feature to be $1$ if it is measured as $1$ at any point
during the tranche in question.

\subsection{Exploratory analysis of density and ages}

\begin{figure}
\centering
\subfloat[Tranche 1]{
\includegraphics[width=0.48\textwidth]{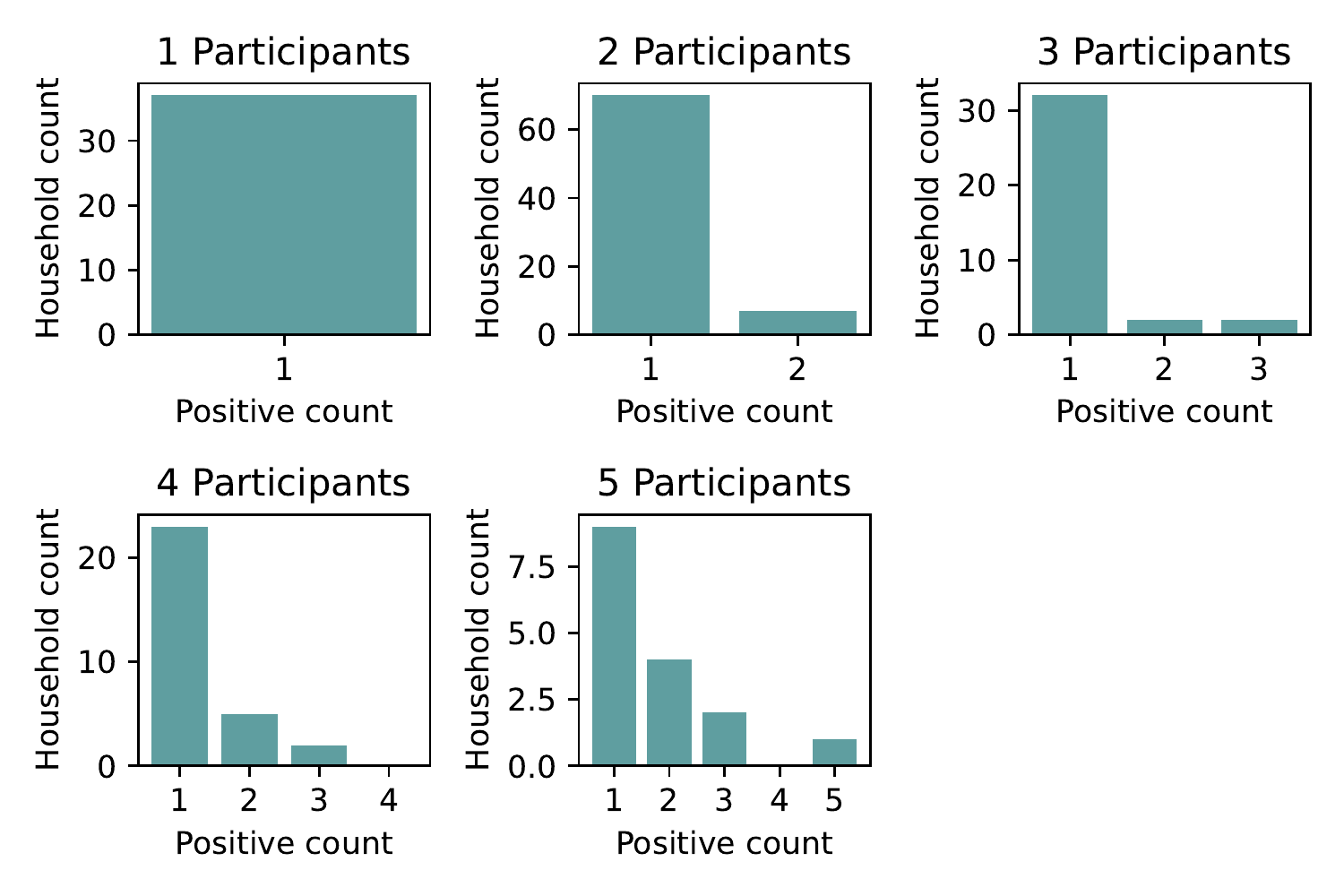}}
\quad
\subfloat[Tranche 2]{
\includegraphics[width=0.48\textwidth]{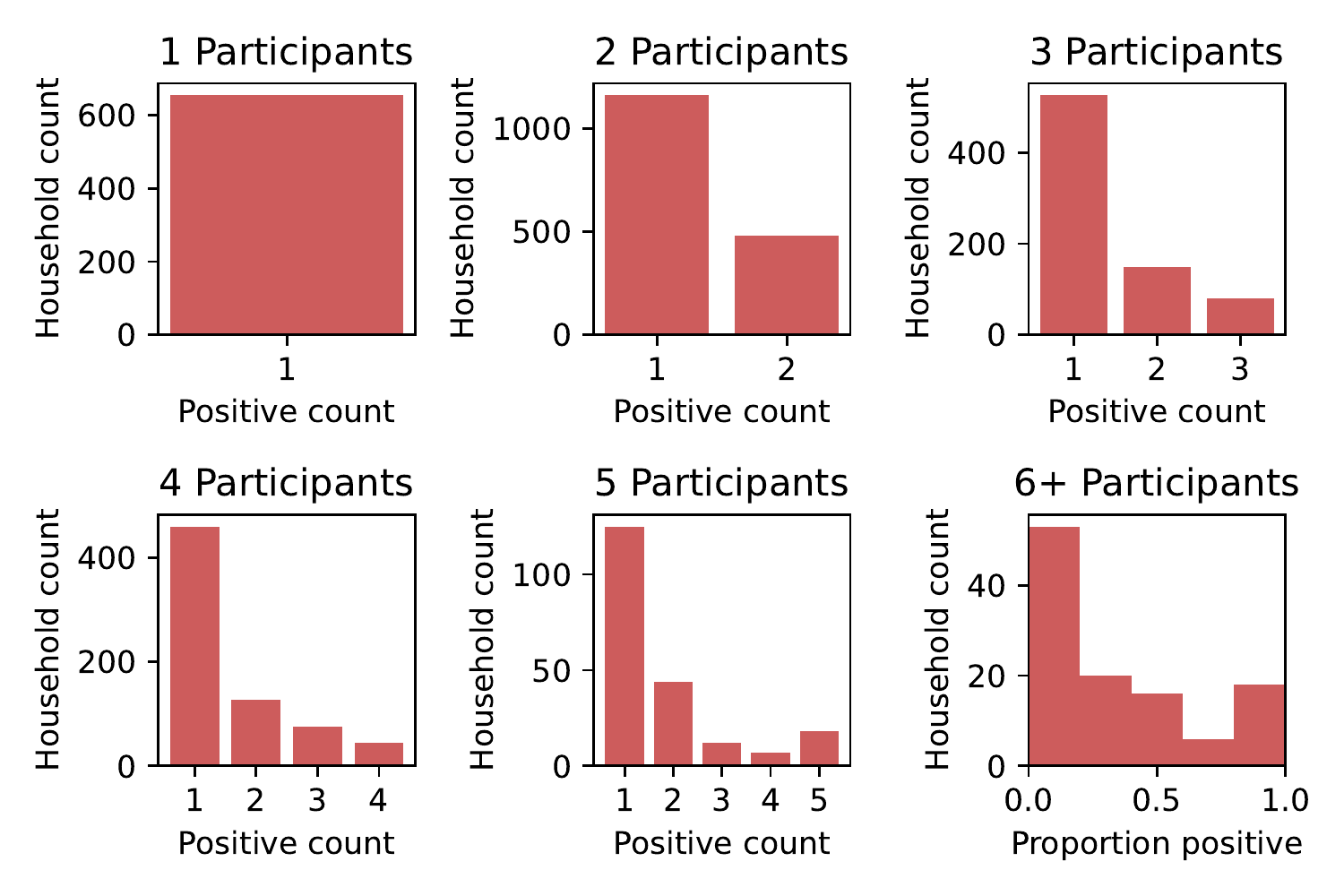}}
\\
\subfloat[Tranche 3]{
\includegraphics[width=0.48\textwidth]{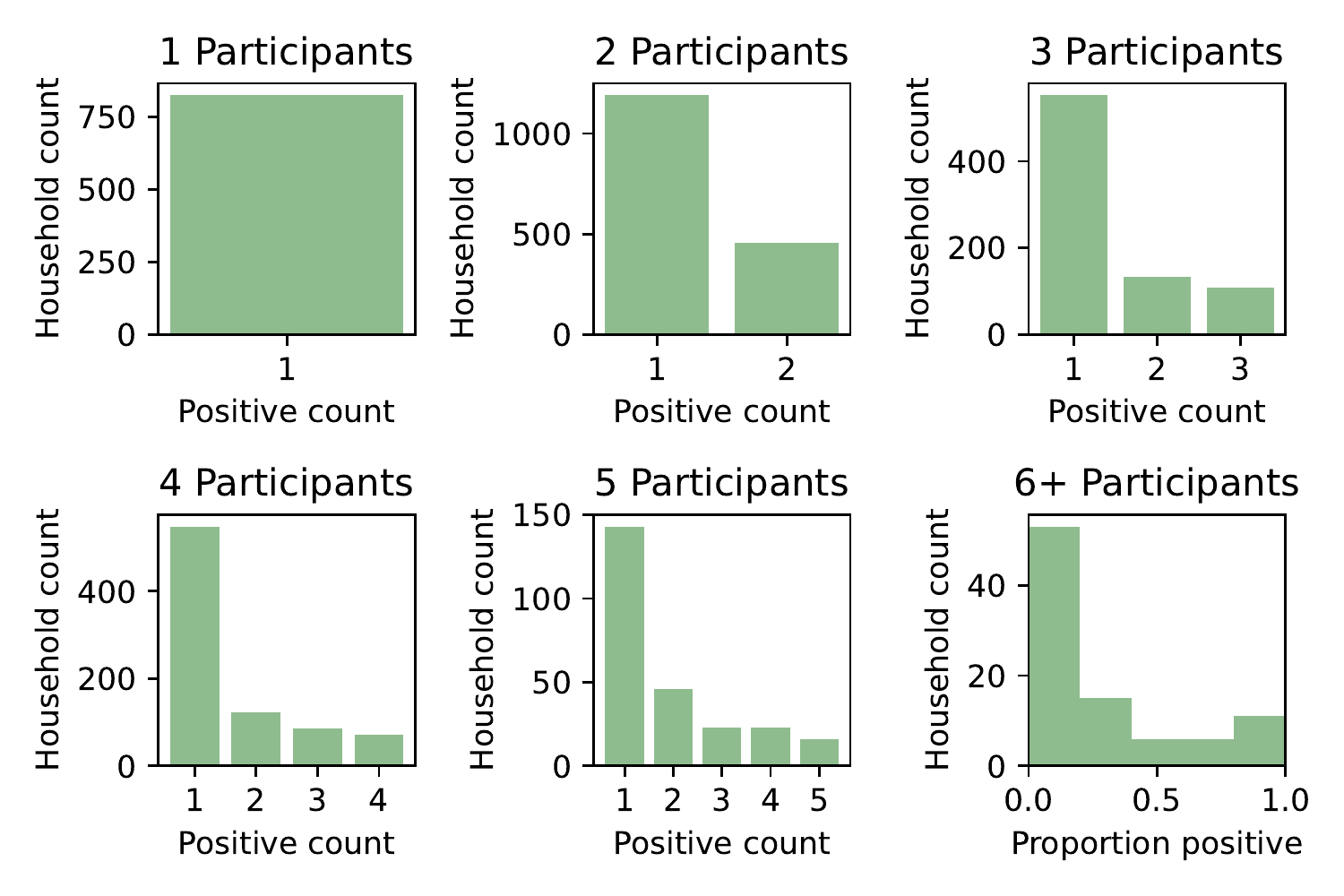}}
\quad
\subfloat[Tranche 4]{
\includegraphics[width=0.48\textwidth]{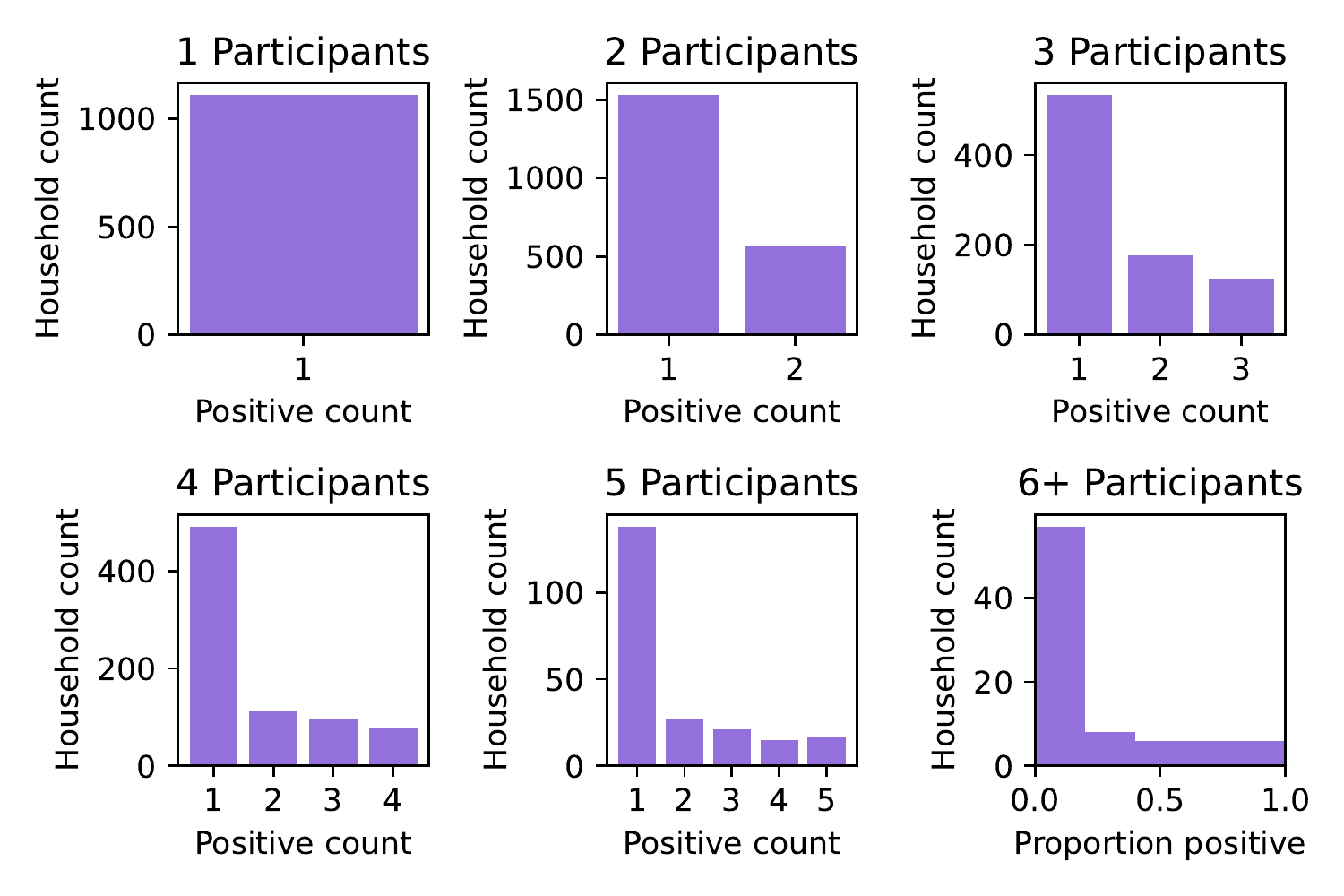}}
\\
\subfloat[Tranche 5]{
\includegraphics[width=0.48\textwidth]{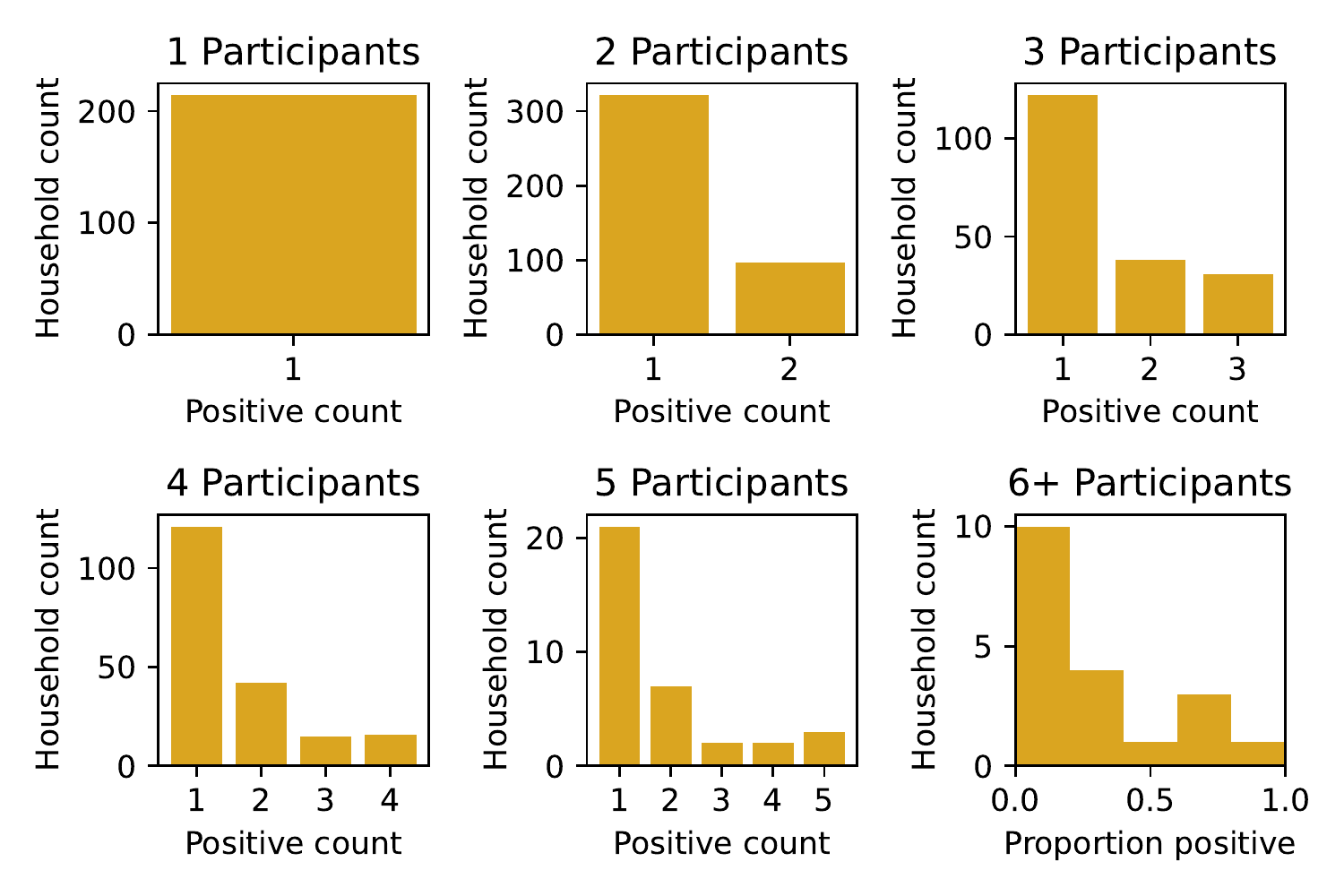}}
\quad
\subfloat[Tranche 6]{
\includegraphics[width=0.48\textwidth]{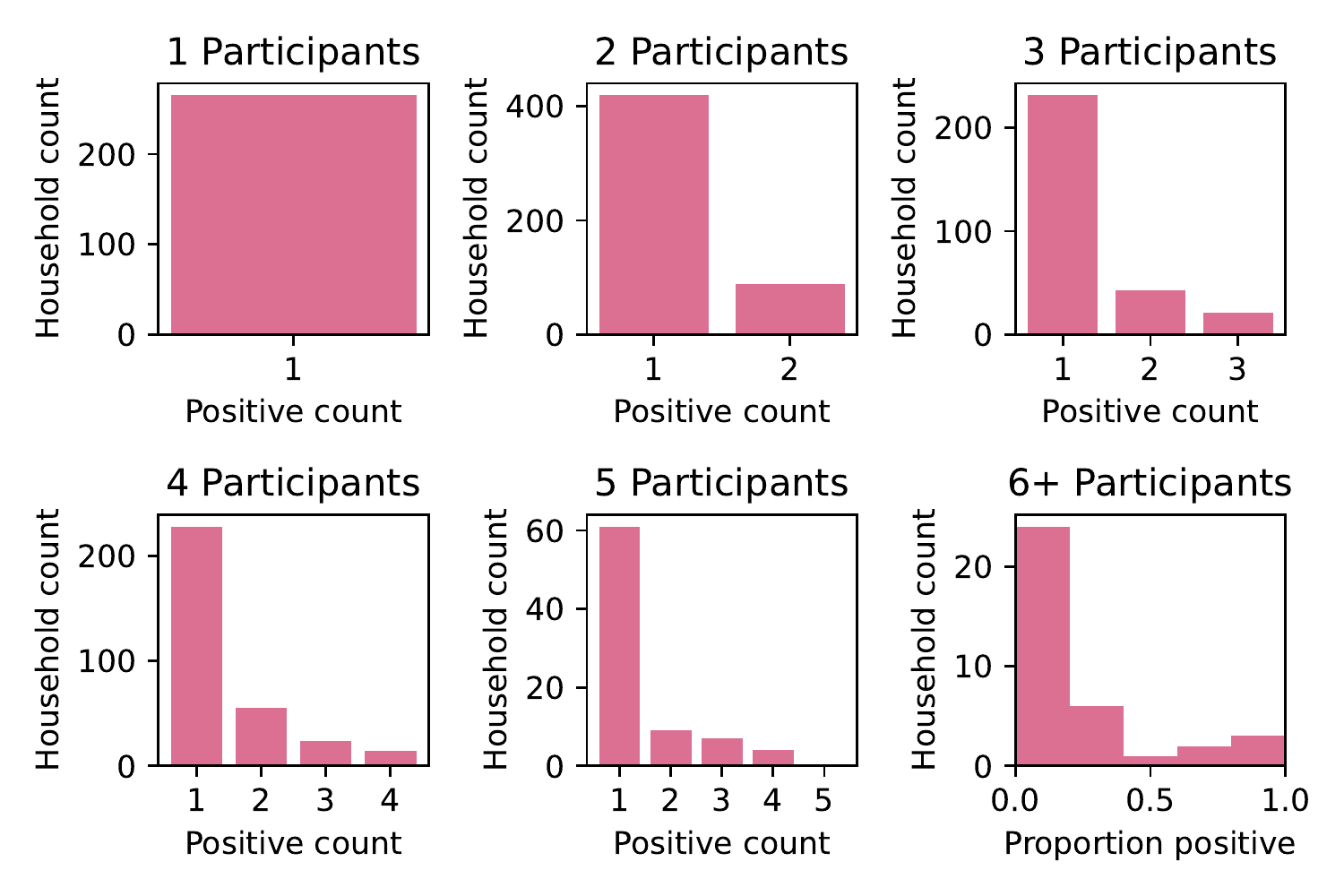}}
\caption{Histograms of household attack rates}
\label{fig:hist}
\end{figure}

\begin{figure}
\centering
\subfloat[Tranche 1]{
\includegraphics[width=0.3\textwidth]{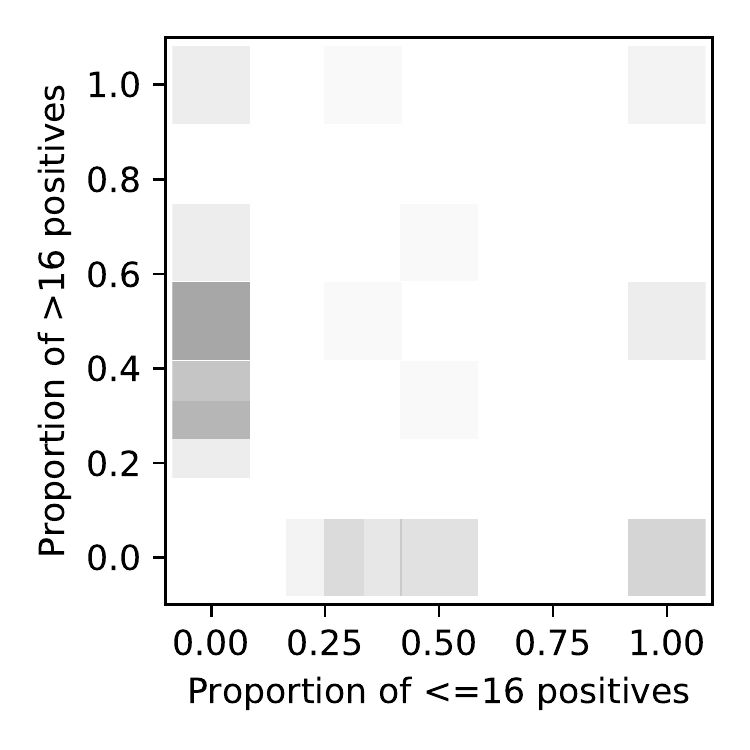}}
\quad
\subfloat[Tranche 2]{
\includegraphics[width=0.3\textwidth]{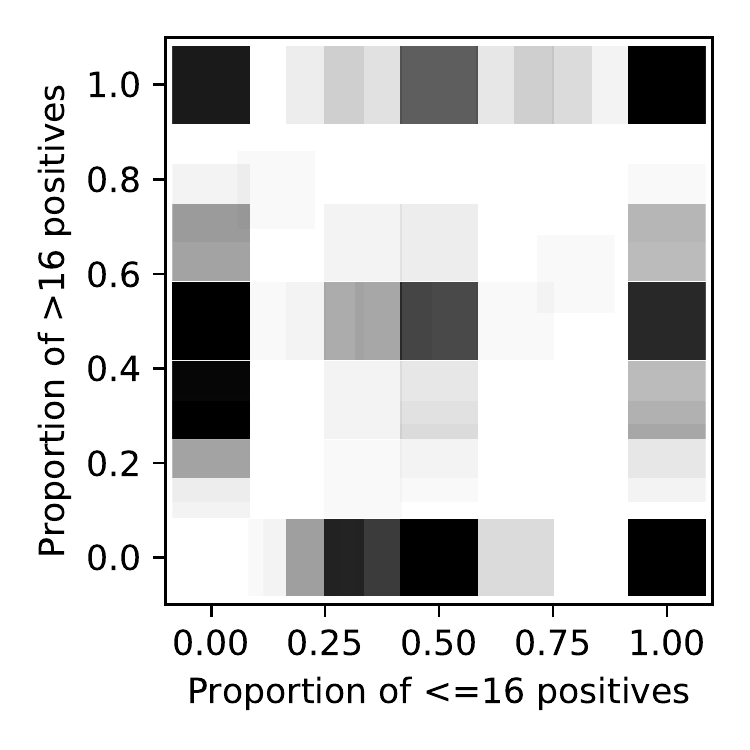}}
\\
\subfloat[Tranche 3]{
\includegraphics[width=0.3\textwidth]{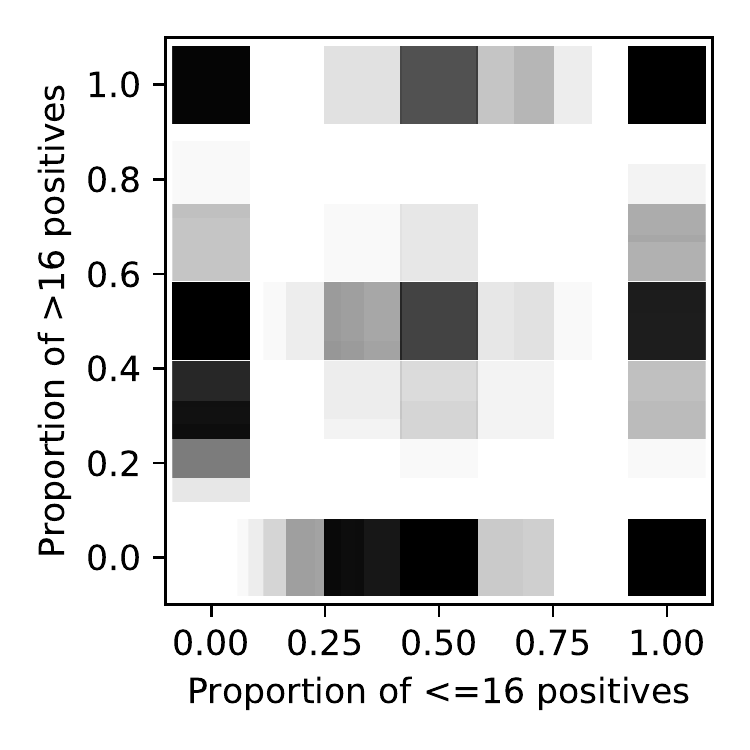}}
\quad
\subfloat[Tranche 4]{
\includegraphics[width=0.3\textwidth]{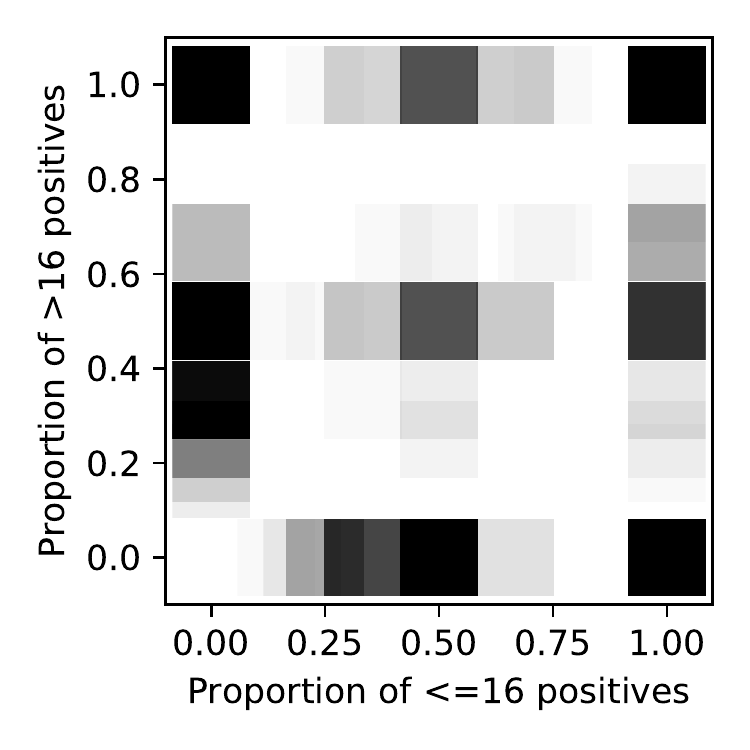}}
\\
\subfloat[Tranche 5]{
\includegraphics[width=0.3\textwidth]{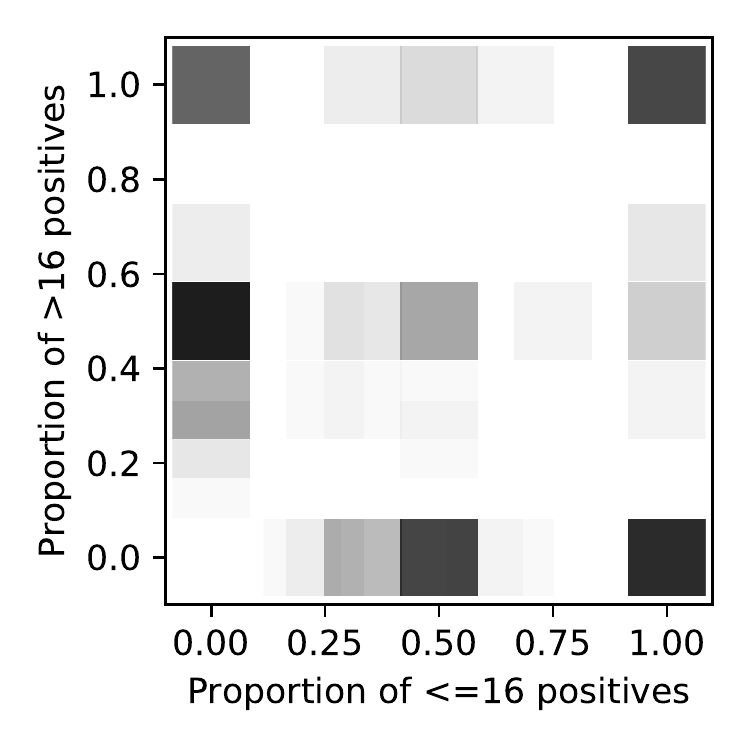}}
\quad
\subfloat[Tranche 6]{
\includegraphics[width=0.3\textwidth]{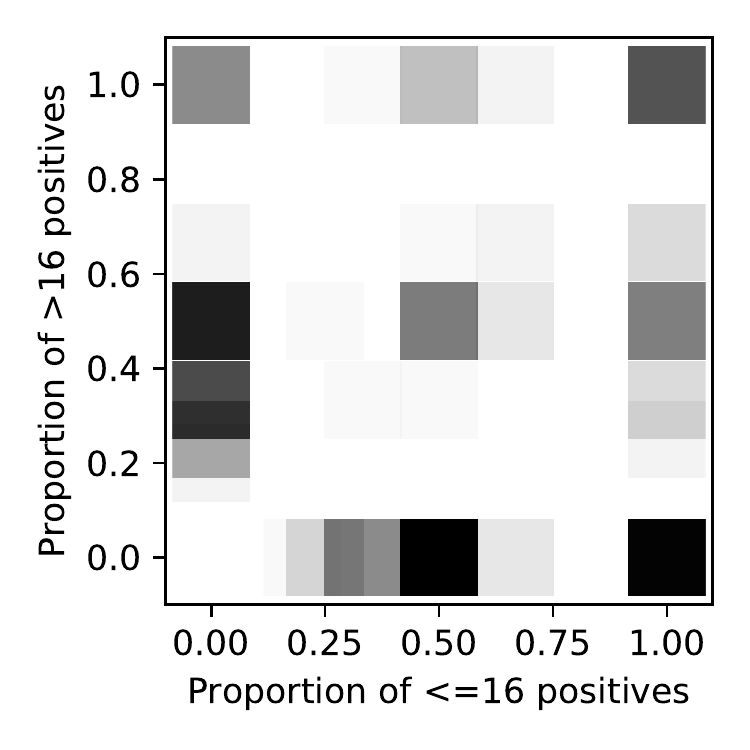}}
\\
\subfloat[Legend]{\includegraphics[width=0.3\textwidth]{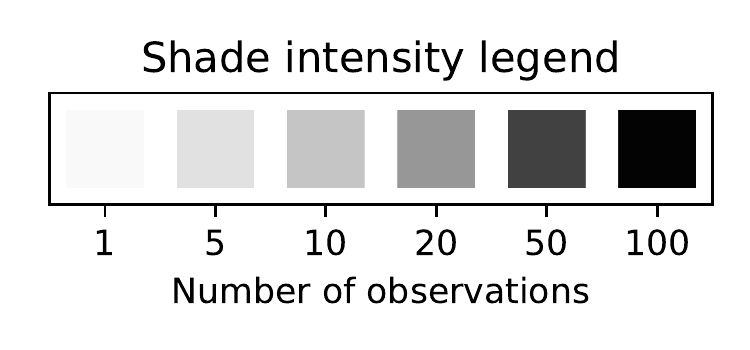}}
\caption{Kernel density plots showing proportion of positives in different age classes
in households.}
\label{fig:chess}
\end{figure}

An important part of our analysis will be consideration of counts / proportions
of households with a given composition of cases displayed as histograms as in
Figure~\ref{fig:hist}, and density plots as in Figure~\ref{fig:chess}.

The heights of the histogram bars are given by 
$$
Z_{k,\ell} = 
\sum_{a=1}^{m} \mathbb{1}_{\{n_a = \ell\}} 
\mathbb{1}_{\{\sum_{i\in H_a} y_i = k\}} \text{ ,}
\quad \ell \in \{ 2,3,4,5,6\} \text{ ,} \quad
k \in\{ 0, \ldots, \ell \} \text{ ,}
$$
where $\mathbb{1}$ stands for the indicator function.  Verbally, $Z_{k,\ell}$
is the count of households of size $\ell$ with $k$ participants testing
positive.

The density plots are obtained by considering some feature (in this case,
age) that takes values $0$ or $1$. We then form a point $\mathbf{r}_a \in
[0,1]^2$ for each household $H_a$ such that
$$
\sum_{i\in H_a} \mathbb{1}_{\{y_i = 1\}} > 0 \text{ ,} \quad
\sum_{i\in H_a} \mathbb{1}_{\{x_i = 1\}} > 0 \text{ ,} \quad
\sum_{i\in H_a} \mathbb{1}_{\{x_i = 0\}} > 0 \text{ ,}
$$
through the definition
$$
\mathbf{r}_a = \left(
\frac{\sum_{i\in H_a}\mathbb{1}_{\{y_i = 1 \& x_i = 1\}}}%
{\sum_{i\in H_a}\mathbb{1}_{\{x_i = 1\}}},
\frac{\sum_{i\in H_a}\mathbb{1}_{\{y_i = 1 \& x_i = 0\}}}%
{\sum_{i\in H_a}\mathbb{1}_{\{x_i = 0\}}}
\right)\text{ .}
$$
Then we can construct a kernel density estimate in the usual way by summing
then normalising kernel functions around the points, in particular the
width-$w$ square kernel function
$$
\mathcal{K}(\mathbf{r},\mathbf{r}_a) = \mathbb{1}_{\{||\mathbf{r} -
\mathbf{r}_a||_{\infty} < w\}} \text{ .}
$$
We use age (16 years old and under versus over 16 years old) as the feature in
making the density plots in Figure~\ref{fig:chess}.

\subsection{Residual analysis and gene positivity pattern}

We are also interested in tabulation of features and positives in households in
a manner that allows their clustering to be assessed.  In particular, this
involves calculation of Pearson residuals for the within-household pairs of
features and positives.  Let $x_i$ be the feature for individual $i$ that takes
values with generic labels $A, B, \ldots$ (here mainly patterns of PCR target
positivity and negativity indicative of viral strain).  We are then interested
in the table of pairs of individuals in households in the set
$\mathcal{H}\subseteq [m]$ with certain properties,
$$
Y_{AB} = \sum_{a\in \mathcal{H}, i\in H_a, j\in H_a \setminus \{i\}}
\mathbb{1}_{\{x_i = A\}} \mathbb{1}_{\{x_j = B\}} \text{ .}
$$
Verbally, $Y_{AB}$ is the count in the sample of $A$-$B$ pairs of distinct
individuals in households from the set of households under consideration. On
its own, this does not indicate whether $A$ and $B$ are more strongly
associated with each other in households than would be expected from their
overall prevalence in the household population. If we let
$$
z_A =  \sum_{a\in \mathcal{H}, i\in H_a} \mathbb{1}_{\{x_i = A\}} \text{ ,}
$$
then under the null hypothesis of independence, $\hat{\pi}_A = z_A /
|\mathcal{H}|$ is the maximum likelihood estimator for the population
probability of being in state $A$ and we can then construct an `expected' table
corresponding to each household pair having independent state with elements
$$
E_{AB} = \hat{\pi}_A \hat{\pi}_B \sum_{a\in \mathcal{H}} n_a(n_a-1) \text{ .}
$$
The Pearson residual associated with the $(A,B)$-th table entry is then
\be
R_{AB} = \frac{Y_{AB}-E_{AB}}{\sqrt{E_{AB}}} \text{ .}
\label{RAB}
\ee
In simpler contexts, such residuals are typically asymptotically standard
normal under the null hypothesis \citep{Bishop:1975}. For our case, this simple
result does not follow straightforwardly, but if we consider a sampled
household $H$, let $X_i$ be the random variable state of the $i$-th household
member, and let
$$
Z_A =  \sum_{i\in H} \mathbb{1}_{\{X_i = A\}} \text{ ,}
$$
then the moment generating function for the random vector $\mathbf{Z} = (Z_A)$
under the assumption of independence will be the multinomial
$$
M_{\mathbf{Z}}(\mathbf{t}) = \left( \sum_{A} \pi_A {\rm e}^{t_A}\right)^{|H|}
\text{ .}
$$
We can then calculate moments of the distribution of pairs through differentiation
of this function, for example
$$
\mathbb{E}\left[Z_A\left(Z_B - \mathbb{1}_{\{A=B\}}\right)\right] = \left.
\frac{\partial^2 M}{\partial t_A \partial t_B}\right|_{\mathbf{t} = \mathbf{0}}
= \pi_A \pi_B n(n-1) \text{ .}
$$
And so we can see that $R_{AB}$ as in \eqref{RAB} will be $0$ where there is no
correlation between states at the household level. While explicit calculation
of $\mathrm{Var}(Z_A(Z_B - \mathbb{1}_{\{A=B\}}))$ to determine its asymptotic
distribution in the case of many households is beyond the scope of the current
work, we believe that this would be an interesting direction for future study.
Nevertheless, due to the arguments presented above we can interpret larger
values of $R_{AB}$ as indicative of more positive correlation between states at
the household level and vice versa.

\begin{figure}
\centering
\subfloat[Tranche 1]{
\includegraphics[width=0.3\textwidth]{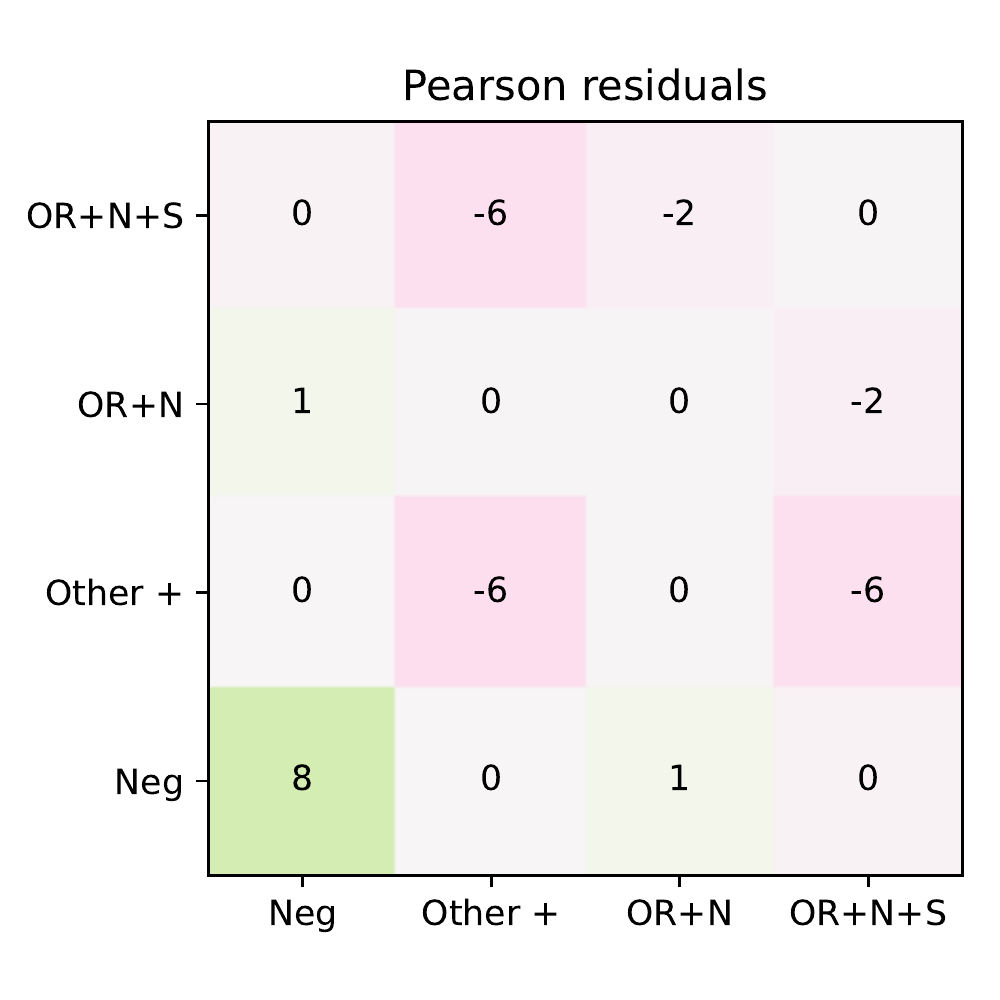}}
\quad
\subfloat[Tranche 2]{
\includegraphics[width=0.3\textwidth]{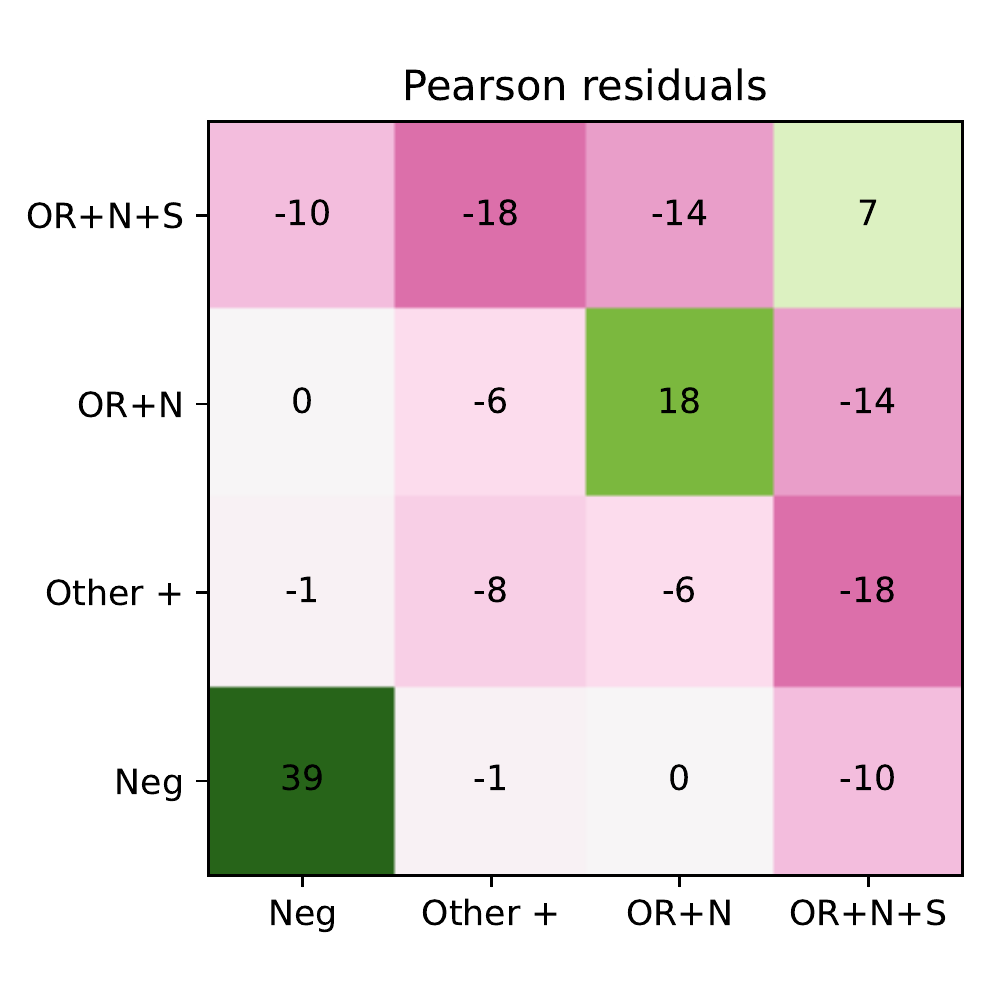}}
\\
\subfloat[Tranche 3]{
\includegraphics[width=0.3\textwidth]{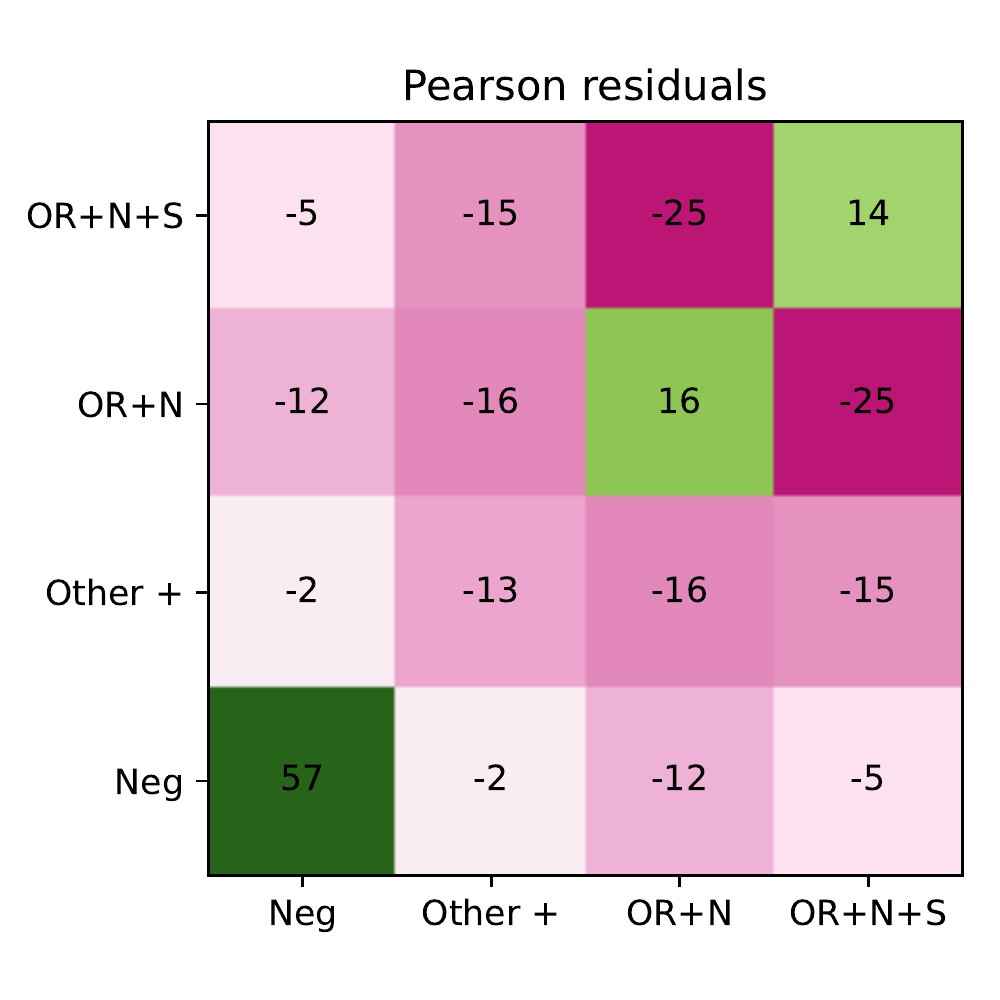}}
\quad
\subfloat[Tranche 4]{
\includegraphics[width=0.3\textwidth]{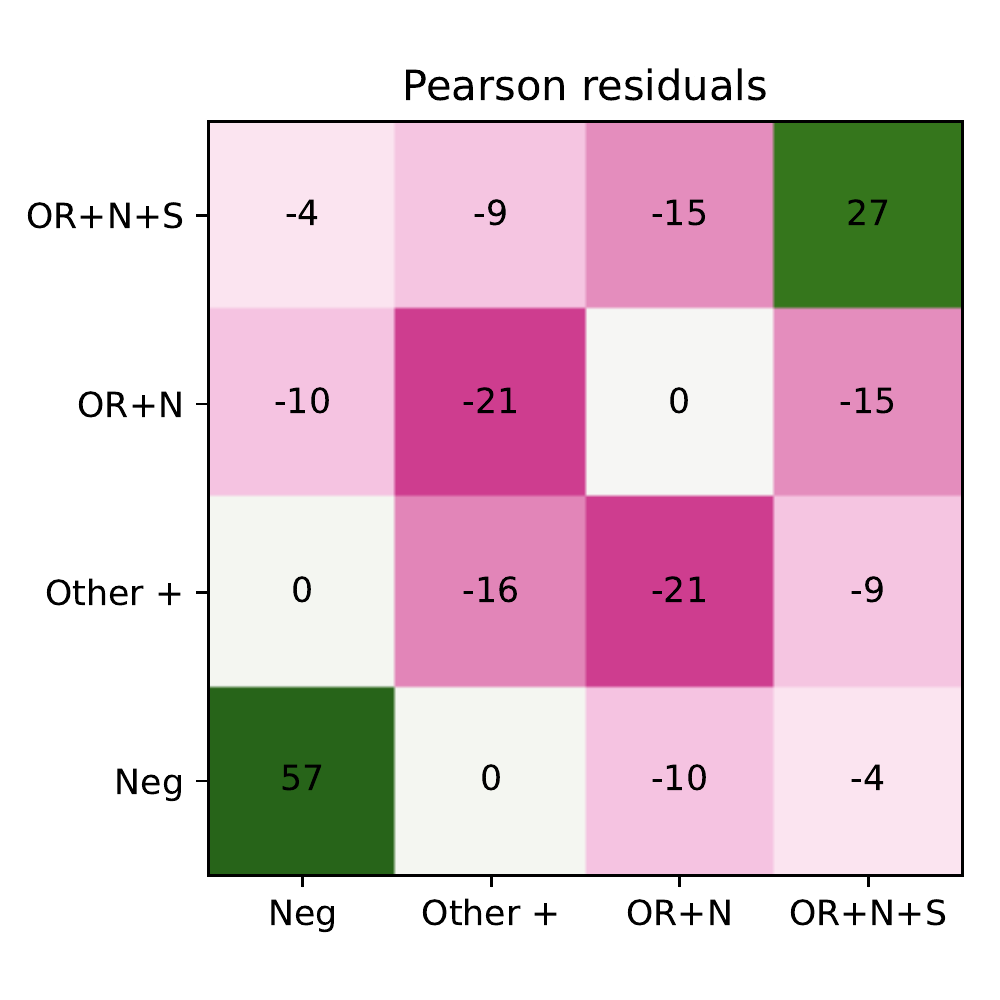}}
\\
\subfloat[Tranche 5]{
\includegraphics[width=0.3\textwidth]{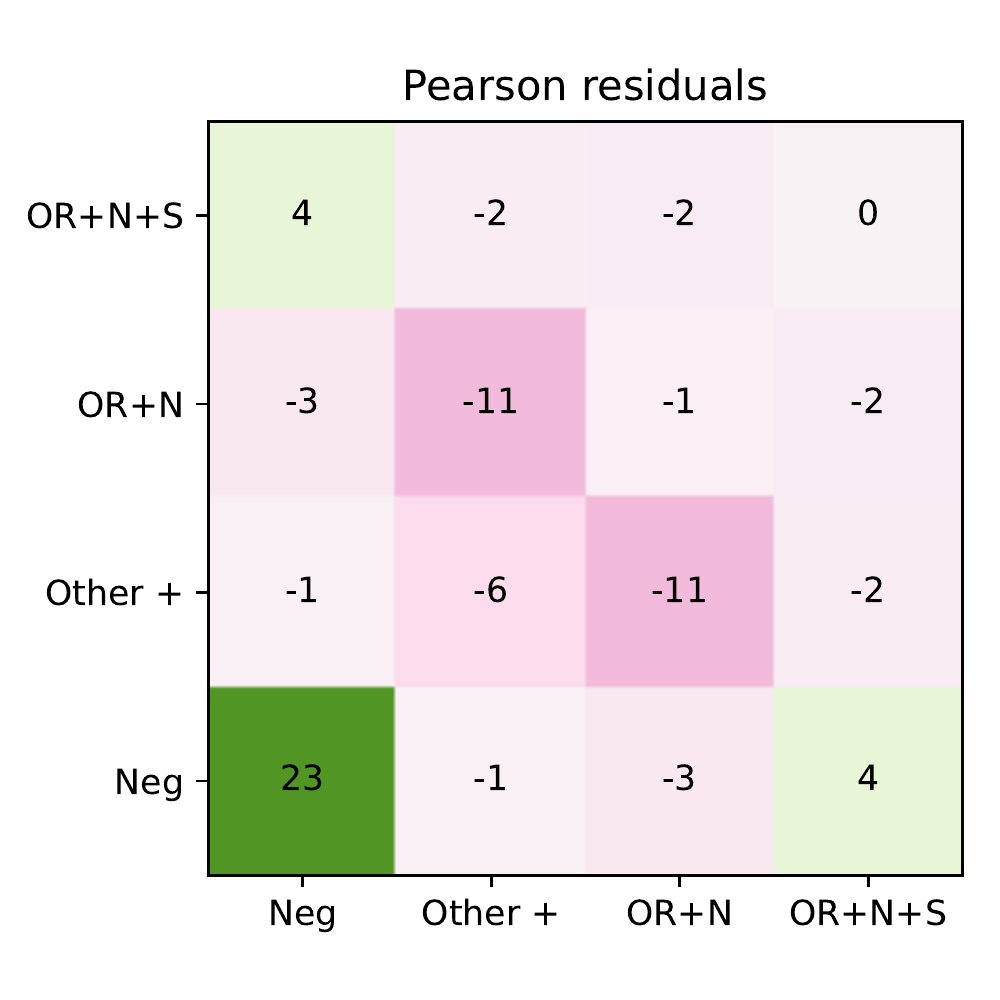}}
\quad
\subfloat[Tranche 6]{
\includegraphics[width=0.3\textwidth]{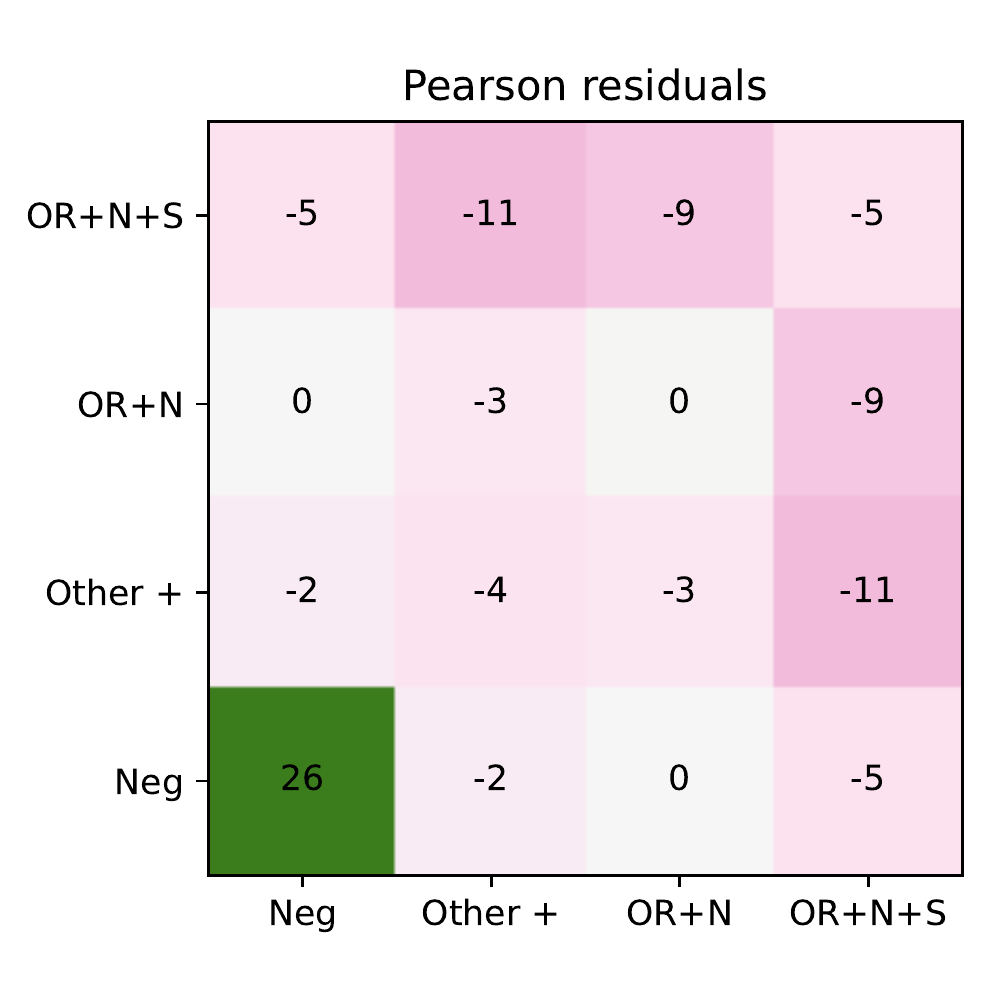}}
\\
\caption{Residual plots for PCR gene positivity patterns.}
\label{fig:resid}
\end{figure}

Here we will use pattern of PCR target failure as a feature and the restriction
of households to those in which there is at least one infection (to avoid
domination of the tables by all-negative households), i.e.\
$$
\mathcal{H} = \bigg\{ a\in[m] \;\bigg|\; \sum_{i\in H_a} y_i > 0 \bigg\} \text{ ,}
$$
to produce the plots in Figure~\ref{fig:resid}.

There are three main patterns of gene positivity that we are interested in:
\textbf{OR+N+S}, which is generally seen in common pre-Alpha variants and the
Delta variant; \textbf{OR+N}, which is associated with the Alpha variant; or
\textbf{Other}, which is usually indicative of too low a viral load to be
confident in strain.  Where an individual is positive on multiple visits with
varying PCR gene positivity patterns, here and throughout we consider the
\textit{maximal} pattern, i.e.\ that containing the least target failures. So
for example, an individual with an N+S positive at one visit followed by an
OR+N+S positive at the next visit and then an N positive at the next visit
would be counted as an OR+N+S positive overall.

\subsection{Full probability model}

While the more exploratory methods above are useful for formulating hypotheses,
the main part of our analysis will be household regression, using time, household
size and individual features to predict positivity. We start by defining
a vector and matrix for each household $H_a, a\in [m]$,
\be
\mathbf{y}_a := (y_i)_{i\in H_a} \text{ ,} \qquad
\boldsymbol{X}_a := [(\mathbf{x}_i)_k]_{i\in H_a, k\in [p]} \text{ .}
\ee
Note that the outcomes of swab positivity are not independent of each other due to
transmission within households, but otherwise the households are selected as
uniformly as possible from the population. This means that an
independent-households assumption is appropriate, in which we write the
likelihood function as
\be
L(\boldsymbol{\theta}) = \prod_{a\in[m]} P_{\mathbf{y}_a}(\boldsymbol{X}_a,
\boldsymbol{\theta}) \text{ .} \label{likeq}
\ee
Here, $\boldsymbol{\theta}$ is a vector of model parameters, and
$P_{\mathbf{y}}$ is a function mapping a household feature matrix and
set of model parameters onto a probability of a given set of household 
positivity outcomes. We can derive a set of equations for such probabilities
from equation (4) of \citet{Addy:1991} as in \citet{Kinyanjui:2019}, and
which we present now with some explanation but not a formal derivation
of all components.

We will consider the relevant equations for a household $H$ of size $n$
with outcome vector $\mathbf{y}$ and feature matrix $\boldsymbol{X}$ (i.e.\
suppressing the household index $a$ to simplify notation). In particular, given
a map $\iota:\{0,1\}^n \rightarrow \{1, \ldots, 2^n\}$, we will be able to form the vector
$\mathbf{P} = (P_{\iota(\mathbf{y})})_{\mathbf{y}\in\{0,1\}^n}$ of
probabilities of different outcomes in the household. This will be a solution
to the set of linear equations
\be
\boldsymbol{B}(\boldsymbol{\theta})\mathbf{P} = \mathbf{1} \text{ ,}
\label{bemat}
\ee
where $\mathbf{1}$ is a length-$2^n$ vector of all ones, and $\boldsymbol{B}
= [B_{\iota(\boldsymbol{\nu}), \iota(\boldsymbol{\omega})}]_{\boldsymbol{\nu},
\boldsymbol{\omega} \in \{0,1\}^n}$, which has 
\be
B_{\iota(\boldsymbol{\nu}), \iota(\boldsymbol{\omega})}
= \mathcal{B}_{\boldsymbol{\nu}, \boldsymbol{\omega}}
= \frac{1}{\prod_{j\in H} \Phi\left( \sum_{i\in H}(1-\nu_i)
\lambda_{ij}\right)^{\omega_j} Q_{j}^{1-\nu_j}} \text{ ,}
\quad 
\boldsymbol{\nu} \leq \boldsymbol{\omega} \in \{0,1\}^n
\text{ ,}
\label{balleq}
\ee
and other elements equal to zero, where we write $\leq$ between vectors to
stand for the statement that each element on the left-hand side is less than or
equal to the corresponding element on the right-hand side. The associated
condition imposes that each $\boldsymbol{\nu}$ above will correspond to a
sub-epidemic of $\boldsymbol{\omega}$ meaning that the equation~\eqref{bemat}
can be solved iteratively. There are then three main ingredients of the
transmission model that we will enumerate below and in doing so define the
terms in Equation~\eqref{balleq}.

The first model component is the probability of avoiding infection from outside;
for the $i$-th individual this is
\be
Q_i = \mathrm{e}^{-\Lambda_i} \text{ ,} \qquad
\Lambda_i = \Lambda \mathrm{e}^{\boldsymbol{\alpha}\cdot \mathbf{x}_i}
= \mathrm{e}^{\alpha_0 + \boldsymbol{\alpha}\cdot \mathbf{x}_i}
\text{ .}
\ee
In the language of infectious disease modelling, $\Lambda_i$ is the cumulative
force of infection experienced by the $i$-th individual. Then $\exp(\alpha_k)$
is the relative external exposure associated with the $k$-th feature /
covariate, meaning that it is the multiplier in front of the baseline force of
infection, which is that for an individual whose feature vector is all zeros,
$\mathbf{0}$. This baseline probability of \emph{avoiding} infection from
outside is then
\be
q = \exp(-\Lambda) = \exp(-\exp(\alpha_0)) \text{ .}
\label{qeq}
\ee
Because this is often much closer to $1$ than to $0$, we will report the
probability of \emph{being} infected from outside the household as a
percentage, i.e.\ $(1-q)\times 100\%$ will be given in figures and tables.  We
will present this alongside the relative external exposures that are elements
of the vector $\boldsymbol{\alpha}$, although it would also be possible to use
\eqref{qeq} to relate $q$ to the baseline force of infection $\Lambda$ or
intercept of the linear predictor, $\alpha_0$.  Note that some care must be
taken in interpretation of this variable when the data are split into time
periods as in this work, since to appear as a household with at least one
positive in one tranche, it is necessary to appear as a household with no
positives in the previous tranches for which the household was in the study.
Values of $1-q$ will typically be low enough here that this conditional
dependence is not strong, but this might not be true at higher levels of
incidence for the same design.

The second component of the model is variability in the infectiousness at the
individual level, usually interpreted as arising from the distribution of
infectious periods. Suppose, in particular, that a household has just one
susceptible and one infectious individual, and that the infectious individual
exerts a force of infection $\lambda$ on the susceptible for a random period of
time $T$. Let the cumulative force of infection be
\be
\label{Ct}
C(t) = \int_{u=0}^{\mathrm{min}(T,t)} \! \! \lambda \, \, \dint{u} \text{ .}
\ee
The first step in analysing this model is to apply the \citet{Sellke:1983}
construction, where the susceptible individual picks a random variable
$\Xi \sim \mathrm{Exp}(1)$ and infection happens once $C(t) > \Xi$, or
no infection happens if $C(T) < \Xi$. To see why this is equivalent to
infection at a rate $\lambda$, take \eqref{Ct} and note that
$$
\mathrm{Pr}(\Xi > C(t+\delta t) |\Xi > C(t) )
= \frac{\int_{0}^{C(t+\delta t)} \mathrm{exp}(-\xi) \dint{\xi}}%
{\int_{0}^{C(t)} \mathrm{exp}(-\xi) \dint{\xi} }
= 1 - \lambda \delta t + o(\delta t) \text{ .} 
$$
The furthest right expression in this equation is what we mean by infection at
a rate.

Using $F_X$ to stand for a cumulative distribution function and $f_X$ for a
probability density function of a random variable $X$, we have the total
probability
of avoiding infection as
\be
\mathrm{Pr}(\Xi > C(T)) = \int_{0}^{\infty} F_{\Xi}(\lambda t) f_{T}(t) \dint{t}
 = \int_{0}^{\infty}\mathrm{e}^{-\lambda t} f_{T}(t) \dint{t}
= \mathcal{L}[f_T](\lambda) =: \Phi(\lambda) \text{ ,} \label{PhiLam}
\ee
where $\mathcal{L}$ stands for Laplace transformation. We can then use this
result to write down the probabilities of different outcomes in a two-person
household without covariates:
$$
\mathrm{Pr}(\mathbf{y}=(0,0)) = Q^2 \text{ ,} \quad
\mathrm{Pr}(\mathbf{y}=(0,1)) = \mathrm{Pr}(\mathbf{y}=(1,0)) =
 Q(1-Q)\Phi(\lambda) \text{ ,}
$$
which are expressions that can also be obtained from \eqref{balleq}. A more
general argument is presented by \citet{Ball:1986} for the full system of
equations, but the expressions above should give some intuition for why these
hold.

For our modelling, we assume that each individual picks an infectious period
from a unit-mean Gamma distribution since the equations are not sensitive to
the mean and this therefore provides a natural one-parameter distribution with
appropriate support.  The Laplace transform of this as used in \eqref{balleq}
is
\be
\Phi(s) = (1+\vartheta s)^{-1/\vartheta} \text{ .} 
\ee
The parameter $\vartheta$ is the variance of the Gamma distribution, i.e.\ it is
larger for more individual variability.

The third component of the model is the infection rate from individual $j$ to
individual $i$,
\be
\lambda_{ij} = n^{\eta} \lambda \sigma_i \tau_j =
n^{\eta} \lambda \mathrm{e}^{\boldsymbol{\beta}\cdot\mathbf{x}_i}
 \mathrm{e}^{\boldsymbol{\gamma}\cdot\mathbf{x}_j}
= \mathrm{e}^{\boldsymbol{\beta}\cdot\mathbf{x}_i}
 \mathrm{e}^{\gamma_0 + \eta\log(n) + \boldsymbol{\gamma}\cdot\mathbf{x}_j}
\text{ .} \label{lij}
\ee
In this equation: $\lambda$ is the baseline rate of infection; $\sigma_i =
\mathrm{e}^{\boldsymbol{\beta}\cdot\mathbf{x}_i}$ is the relative
susceptibility of the $i$-th participant, and $\exp(\beta_k)$ is the relative
susceptibility associated with the $k$-th feature; $\tau_j =
\mathrm{e}^{\boldsymbol{\gamma}\cdot\mathbf{x}_j}$ is the relative
transmissibility of the $j$-th participant, and $\exp(\gamma_k)$ is the
relative transmissibility associated with the $k$-th feature / covariate. As
can be seen from \eqref{lij}, we can interpret $\log(\lambda)$ as the intercept
of the linear predictor for transmissibility.  The term $n^{\eta}$ is a
modelling approach to the effect of household size usually attributed to
\citet{Cauchemez:2004}; as can be seen from \eqref{lij}, this is equivalent to
taking $\log(n)$ as a covariate for transmissibility. Experience with fitting
these models \citep{Kinyanjui:2018} suggests that it is a good idea to impose
hard bounds on the Cauchemez parameter, i.e.\ insist that $\eta \in
[\eta_{\mathrm{min}}, \eta_{\mathrm{max}}]$, meaning that here we will treat
$\eta$ separately from other parameters.

\subsection{Model variables and fitting}

We now enumerate all of the model parameters, distinguishing between the
`natural' representations of parameters that sit in $\mathbb{R}$ and transforms
of natural parameter space $\mathbb{R}^{\kappa}$ that are most
epidemiologically interpretable and therefore suitable for reporting. 
Since $\Lambda$, $\lambda$ and $\vartheta$ have positive support, we can 
use logarithmic and exponential functions to transform between epidemiological
and natural parameters. As noted above, we want $\eta$ to have compact support,
and so note that the function $\mathrm{tan}:[-\pi/2,\pi/2] \rightarrow
\mathbb{R}$ and its inverse can be used. We choose $\eta_{\mathrm{min}} =
-2$ and $\eta_{\mathrm{max}} = 2$, meaning that our natural parameter vector is
$\boldsymbol{\theta} = (\log(\Lambda), \log(\lambda), \log(\vartheta), \tan(\pi
\eta/4), \boldsymbol{\alpha}, \boldsymbol{\beta}, \boldsymbol{\gamma}) \in
\mathbb{R}^{\kappa}$.  

The first part of this parameter vector is the external force of infection,
with natural representation $\alpha_0 =\log(\Lambda)$. Here we will quote the
baseline probability of infection from outside as a percentage, which is
$(1-q)\times 100\%$ for $q$ as in \eqref{qeq}.

The second part of the parameter space relates to baseline within-household
transmission with natural representation $\gamma_0 = \log(\lambda)$,
$\log(\vartheta)$, and $\tan(\pi \eta/4)$, where we use this transform for
$\eta$ to make a hard constraint of epidemiologically meaningful values.  For
interpretability, we work with probabilities of infection by household size,
which from generalising \eqref{PhiLam} to a size-scaled transmission rate are
\be
p_n = 1 - \Phi(n^{\eta} \lambda) \text{ .}
\ee
Such quantities have been called Susceptible-Infectious Transmission Probabilities
(SITP) by e.g.\ \citet{Fraser:2011}, who estimated values close to 20\% from
historical data on the 1918 influenza pandemic.

The third part are features, where we consider:
\begin{itemize}
\item Three age groups: 2-11 years old; 12-16 years old; and older.
\item Working in a patient-facing role or not.
\item Pattern of PCR gene target positivity: OR+N+S, which is associated with
pre-Alpha variants and the Delta variant; OR+N, which is associated with the
Alpha variant; or other, which is usually indicative of too low a viral load to
be confident in strain.
\end{itemize}
We assume that age and working in a patient-facing role have an association
with external risk, leading to natural parameters $\alpha_{\text{2-11}}$,
$\alpha_{\text{12-16}}$ and $\alpha_{\text{PF}}$; that age has an association
with susceptibility, leading to natural parameters $\beta_{\text{2-11}}$ and
$\beta_{\text{12-16}}$; and that age and gene positivity in PCR have an
association with transmissibility, leading to natural parameters
$\gamma_{\text{2-11}}$, $\gamma_{\text{12-16}}$, $\gamma_{\text{OR+N}}$ and
$\gamma_{\text{oth}}$. For any natural parameter $r$, we will report the
multiplicative effect $\exp(r)$.

Model fitting was performed in an approximate Bayesian framework using the
Laplace approximation.  As noted above, households of size 7 and larger were
excluded from the analysis partly because these are often very different in
composition from smallerhouseholds, and partly because of the numerical cost of
solving a linear system of size $2^{n}$ We combine the likelihood
\eqref{likeq} with a standard normal prior on natural parameters,
$\boldsymbol{\theta}\sim\mathcal{N}_{\kappa}(\mathbf{0},\boldsymbol{I})$.
Sensitivity of results to this prior was considered for different variances and
revealed essentially no impact on the highly identifiable parameters such as
$\Lambda$ and $\lambda$, and that while a higher variance could slightly reduce
the shrinkage of effect sizes towards zero, it could also lead to instability
in fitting, meaning that this prior achieves regularisation of the inference
problem without excessive bias.  The maximum a posteriori estimate was obtained
using multiple restarts of a Quasi-Newton optimiser. The Hessian was calculated
numerically for the natural parameters and used in the Laplace approximation to
the posterior on the natural parameters. The credible intervals (CIs) are then
transformed from natural to epidemiologically interpretable parameters.

\subsection{Data processing and software implementation}

The analysis was carried out on the ONS Secure Research Server in the Python 3
language. To illustrate issues with data processing, note that the `flat' form
for the data extracted from the database after cleaning takes a form like:
\begin{verbatim}
    HID   PID   Visit Date   Age   Test Result   Work PF   Pattern
    ...
    123   456   2020-10-02   8     Negative      No        NA
    123   457   2020-10-02   38    Negative      No        NA
    123   456   2020-10-10   8     Negative      No        NA
    123   457   2020-10-10   38    Positive      No        OR+N+S
    123   456   2020-10-17   9     Positive      No        OR+N+S
    123   457   2020-10-17   38    Negative      No        NA
    ...
    124   458   2021-02-15   53    Negative      Yes       NA
    ...
\end{verbatim}
In particular, there is a hierarchical structure to the data. Households, each
with a unique household ID in the \texttt{HID} column, have a number of study
participants with a unique participant ID in the \texttt{PID} column, and each
participant being visited on a number of dates as in the \texttt{Visit Date}
column. Each visit will have associated participant features (e.g.\ as in the
\texttt{Age} column above) and a \texttt{Test Result}.

The large size of this flat file (slightly under three million rows)
means that it is advantageous to use specialist libraries, in this case
\textit{pandas} \citep{reback2020pandas,mckinney-proc-scipy-2010} together with
\textit{NumPy} \citep{harris2020array}. To deal with the nested structure of
the data, we used the `split-apply-combine' paradigm that this library
encourages by analogy with SQL operations. In the example above, this would
involve first associating each participant with an age using
\texttt{pandas.groupby('PID')} and \texttt{pandas.DataFrame.apply(numpy.min)},
and then producing an array of ages for each household using
\texttt{pandas.groupby('HID')} and
\texttt{pandas.DataFrame.apply(numpy.array)}. A similar approach is possible
for test results and multiple features.

Apart from data processing, the main computational cost of the analysis is the
linear algebra associated with solving~\eqref{bemat}, particularly for larger
households. Due to portability, this was carried out in NumPy on the ONS
system, however we found that implementation in \textit{Numba} \citep{Lam:2015}
can generate significant speed-ups, as might use of GPU hardware through use of
e.g.\ \textit{PyTorch} \citep{Pytorch:2015}.

Access to ONS CIS data is possible via the Office for National Statistics'
Secure Research Service, and Python code demonstrating the methodology applied
to publicly available data is at
\url{https://github.com/thomasallanhouse/covid19-housefs}.

\section{Results and Discussion}

\subsection{Exploratory analysis}

Figure~\ref{fig:hist} shows the distribution of positives in households;
comparison with Table~\ref{tab:features} shows that the numbers of households
with two or more positives are much greater than would be expected under the
assumption of independence. In fact, some histograms even take a bimodal `U'
shape.

This multi-modality is even more apparent in the kernel plots in
Figure~\ref{fig:chess}, which also demonstrate that it is common to see
households with only child positives, only adult positives, or both. In
particular, this suggests that both children and adults can be responsible for
bringing infection into the household.  While some of the child-infection-only
households could arise due to failure of ascertainment of an adult infection in
the household, this is unlikely to be true for most, meaning, the introduction
of infection to the household would have been due to a child (and vice versa
for adult-infection-only households).

\subsection{Residual analysis}

\begin{figure}
\centering
\subfloat[Tranche 1]{
\includegraphics[width=0.3\textwidth]{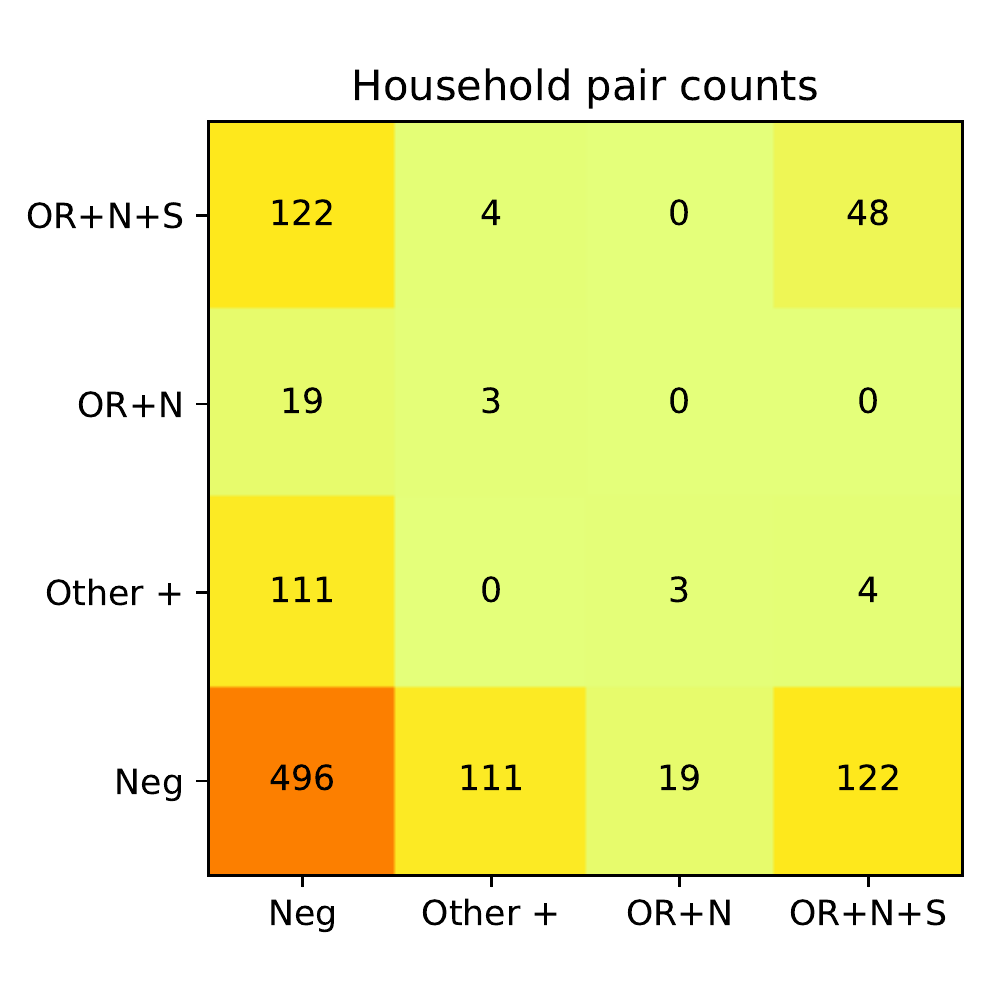}}
\quad
\subfloat[Tranche 2]{
\includegraphics[width=0.3\textwidth]{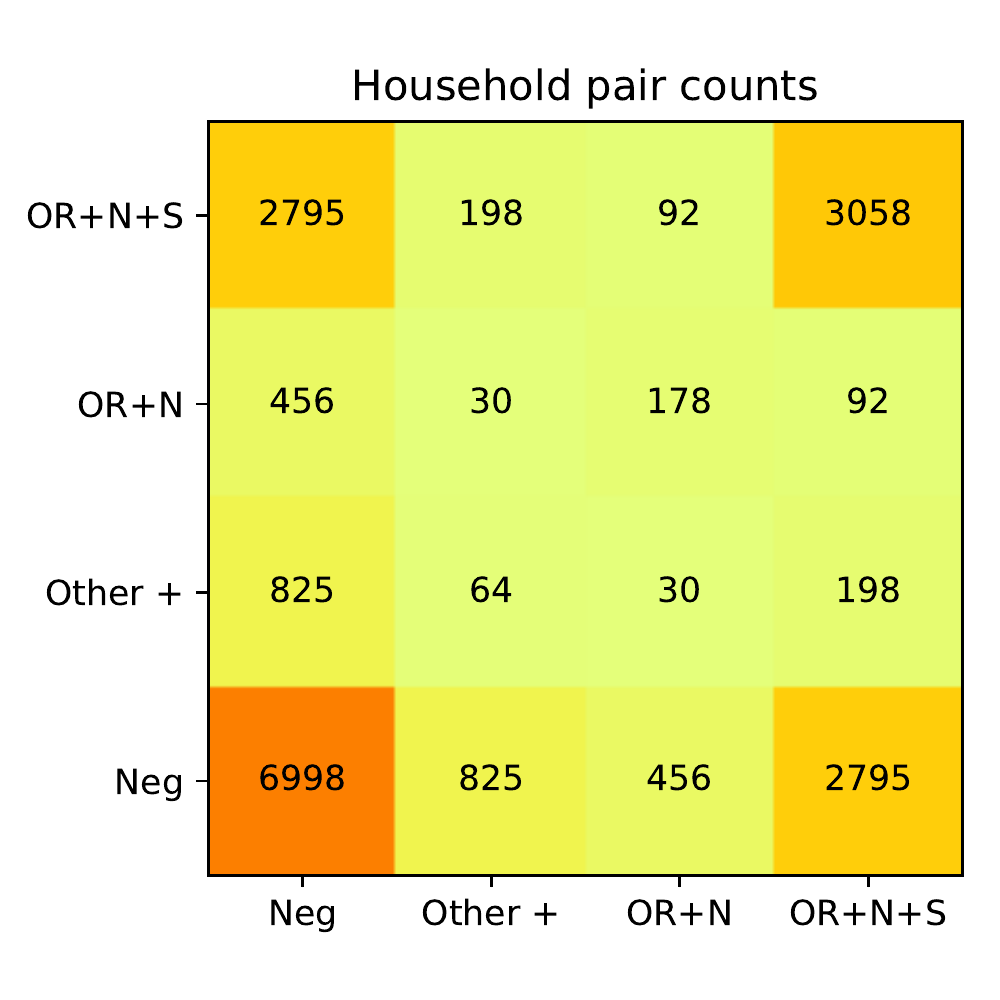}}
\\
\subfloat[Tranche 3]{
\includegraphics[width=0.3\textwidth]{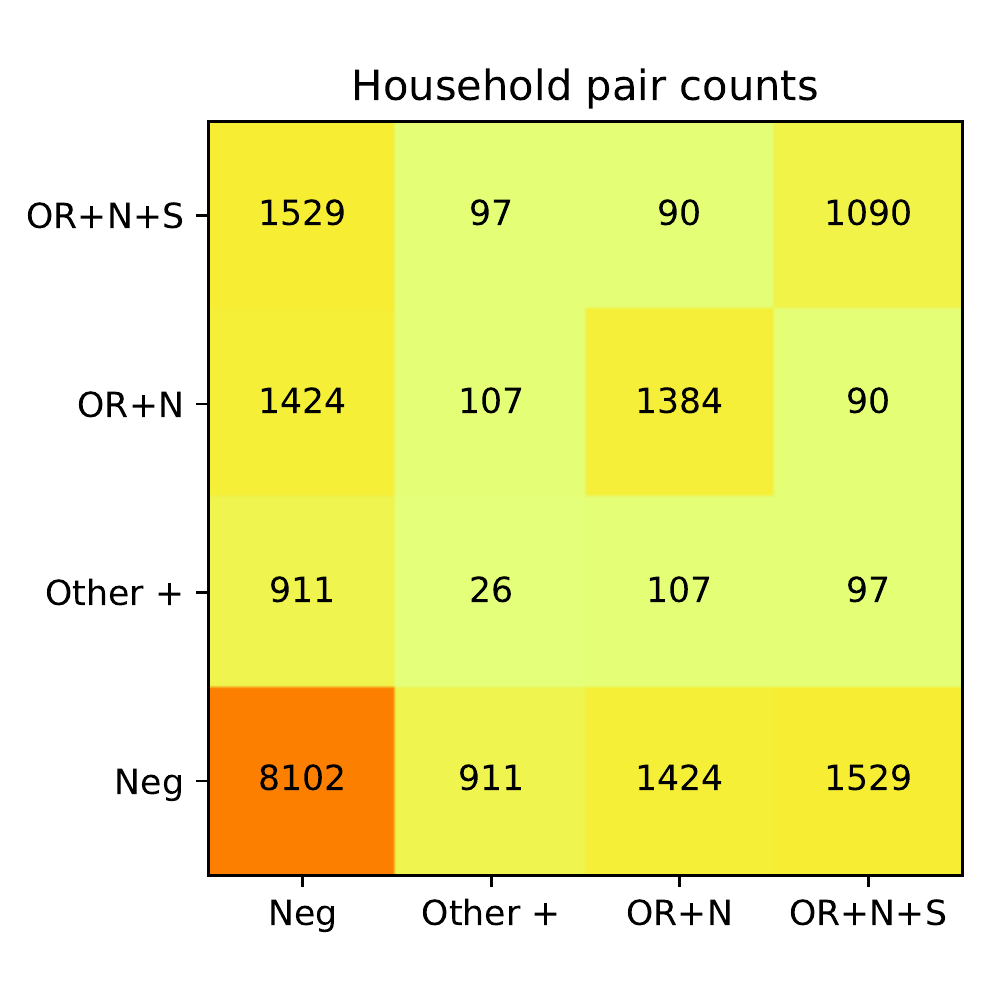}}
\quad
\subfloat[Tranche 4]{
\includegraphics[width=0.3\textwidth]{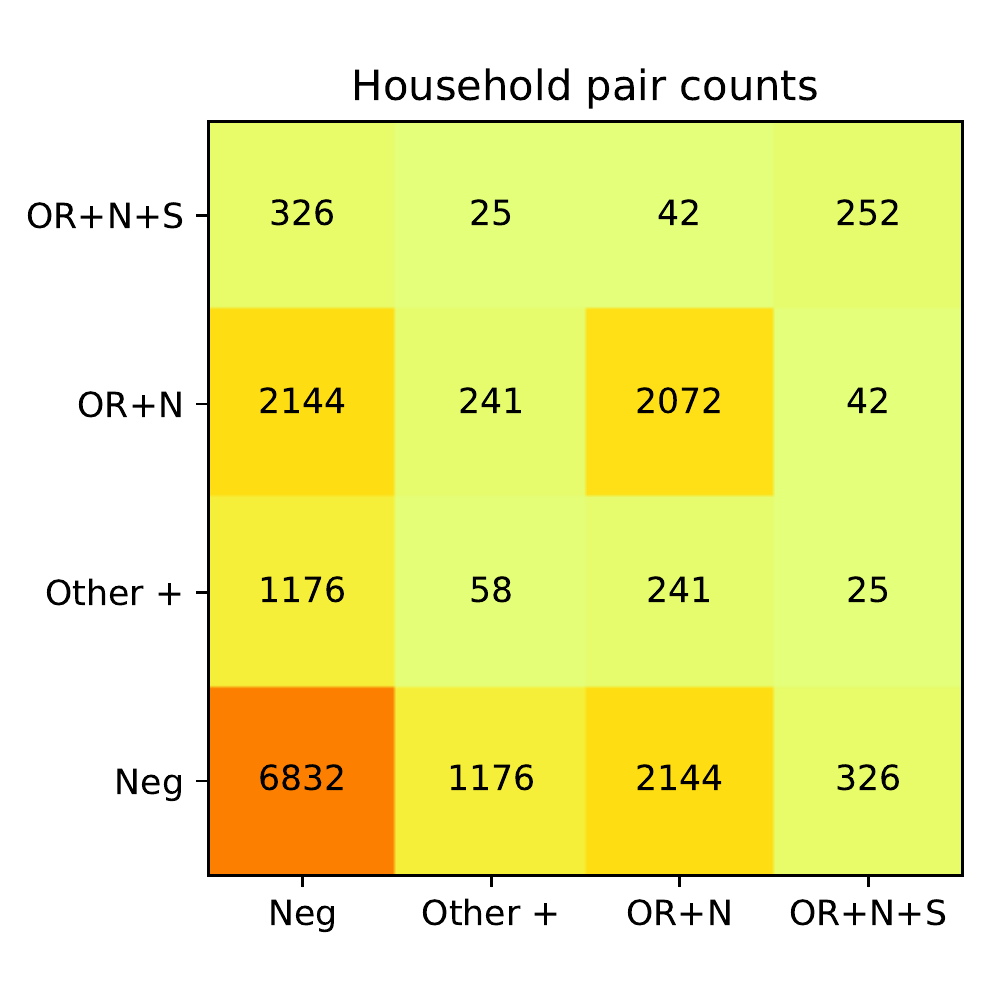}}
\\
\subfloat[Tranche 5]{
\includegraphics[width=0.3\textwidth]{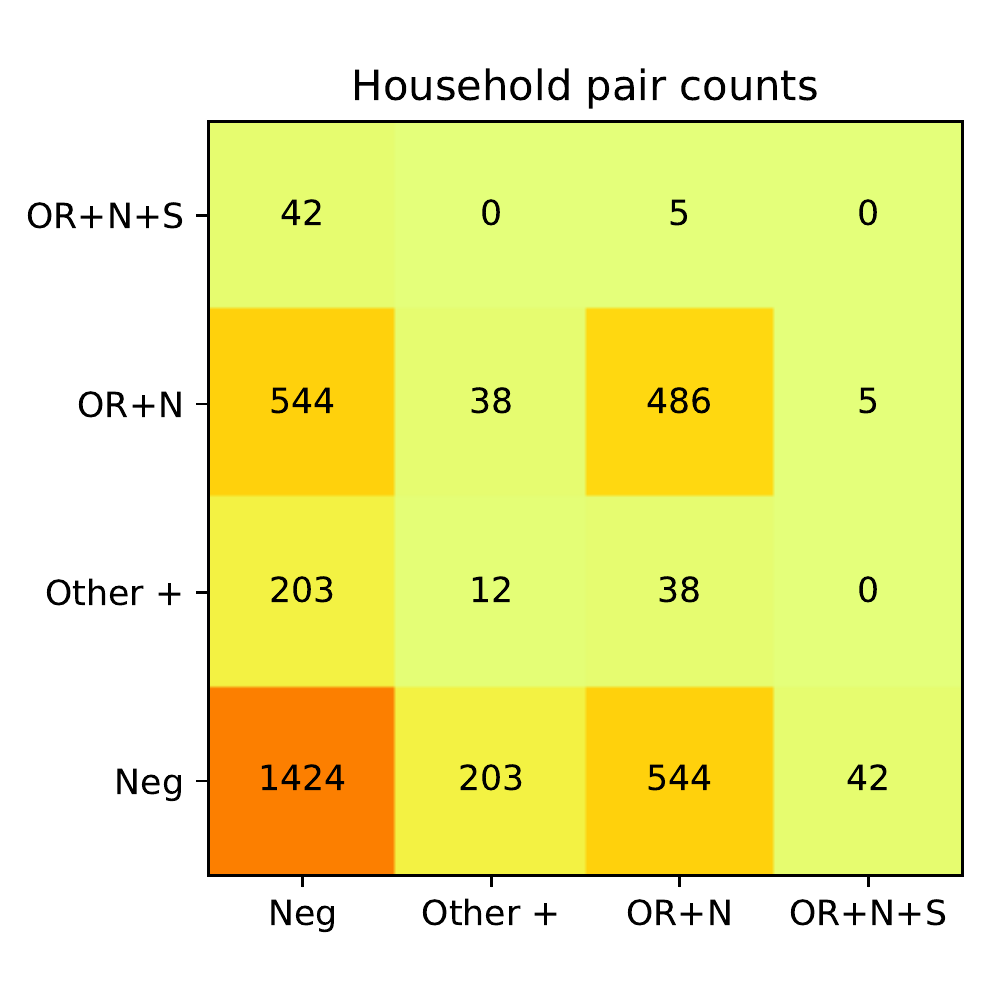}}
\quad
\subfloat[Tranche 6]{
\includegraphics[width=0.3\textwidth]{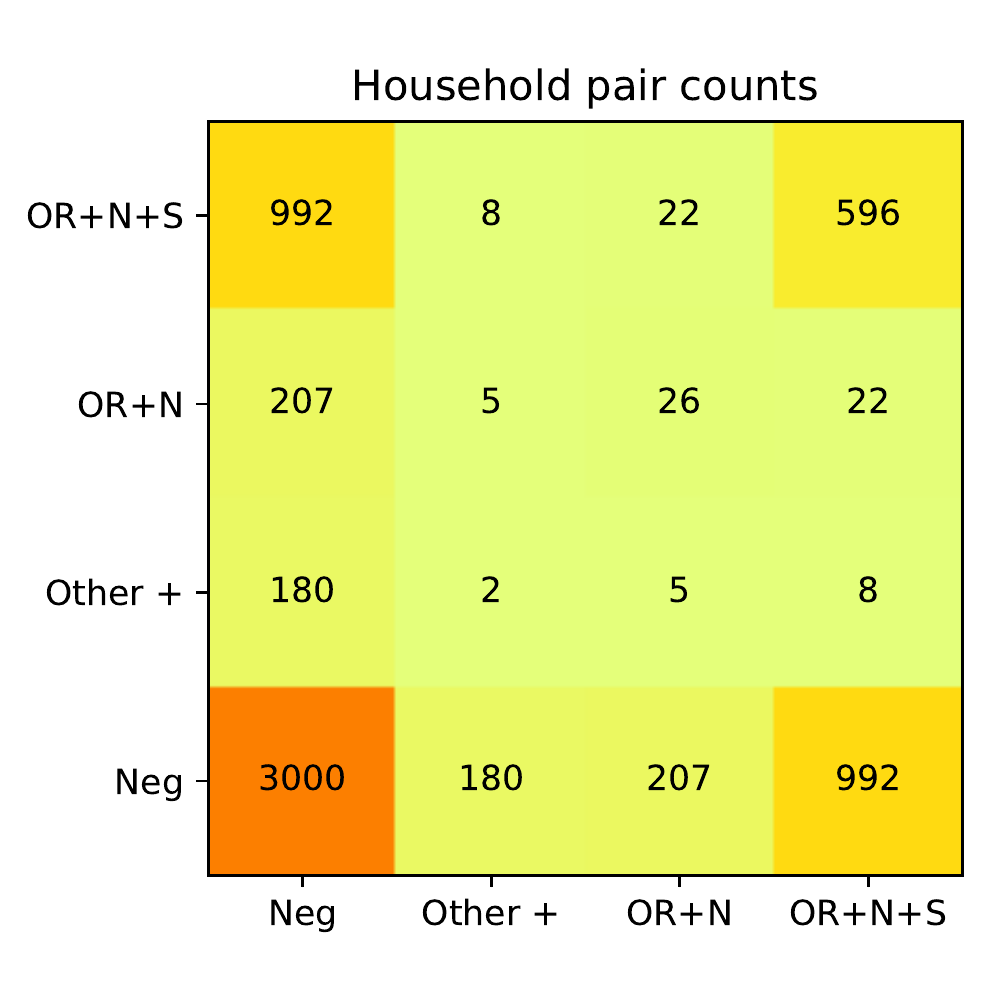}}
\\
\caption{Pair counts for PCR gene positivity patterns.}
\label{fig:pair}
\end{figure}

The pair counts and Pearson residual analysis -- applied to the maximal PCR
target gene positivity pattern being OR+N+S, OR+N, other positive, or negative
-- are shown in Figures~\ref{fig:pair} and~\ref{fig:resid}. The pair counts
show at the household level the replacement of the OR+N+S pattern as the main
source of positive pairs in households with the OR+N pattern, and then the
return of the OR+N+S pattern. We also see from the residual plots that while
there is positive correlation of (OR+N+S)-(OR+N+S) and (OR+N)-(OR+N) pairs, as
well as of negative-negative pairs, there is a negative correlation associated
with (OR+N+S)-(OR+N) pairs and also between pairs involving any other positive
pattern. While this analysis is not mechanistic or causal, we expect that the
main factor generating correlation / clustering of positives in households is
transmission.  As such, the results are consistent with our understanding of
the sweeps of the Alpha and Delta variants as arising due to these being more
transmissible strains than those that they replaced.

\subsection{Regression analysis}

\begin{figure}
\begin{center}
\includegraphics[width=0.6\textwidth]{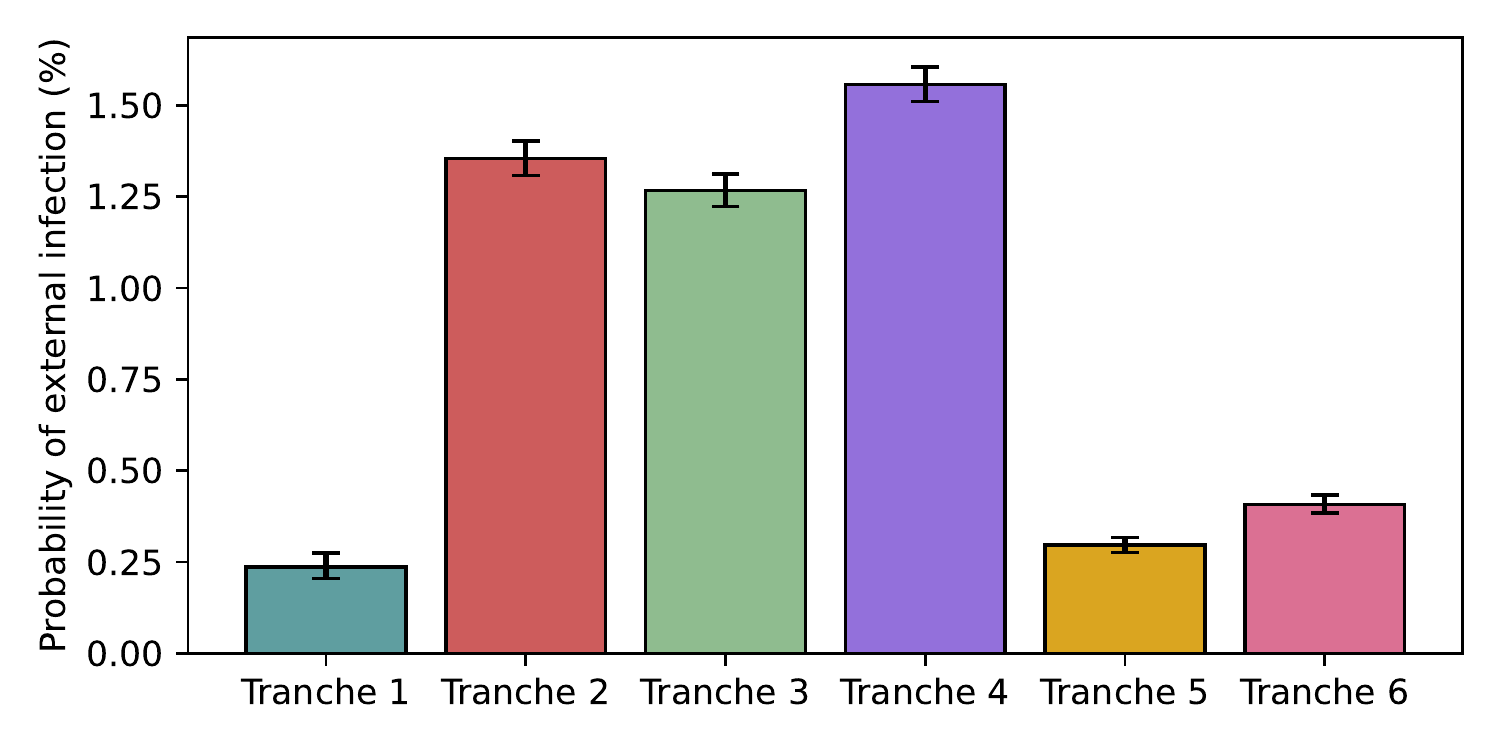}\\
\includegraphics[width=0.8\textwidth]{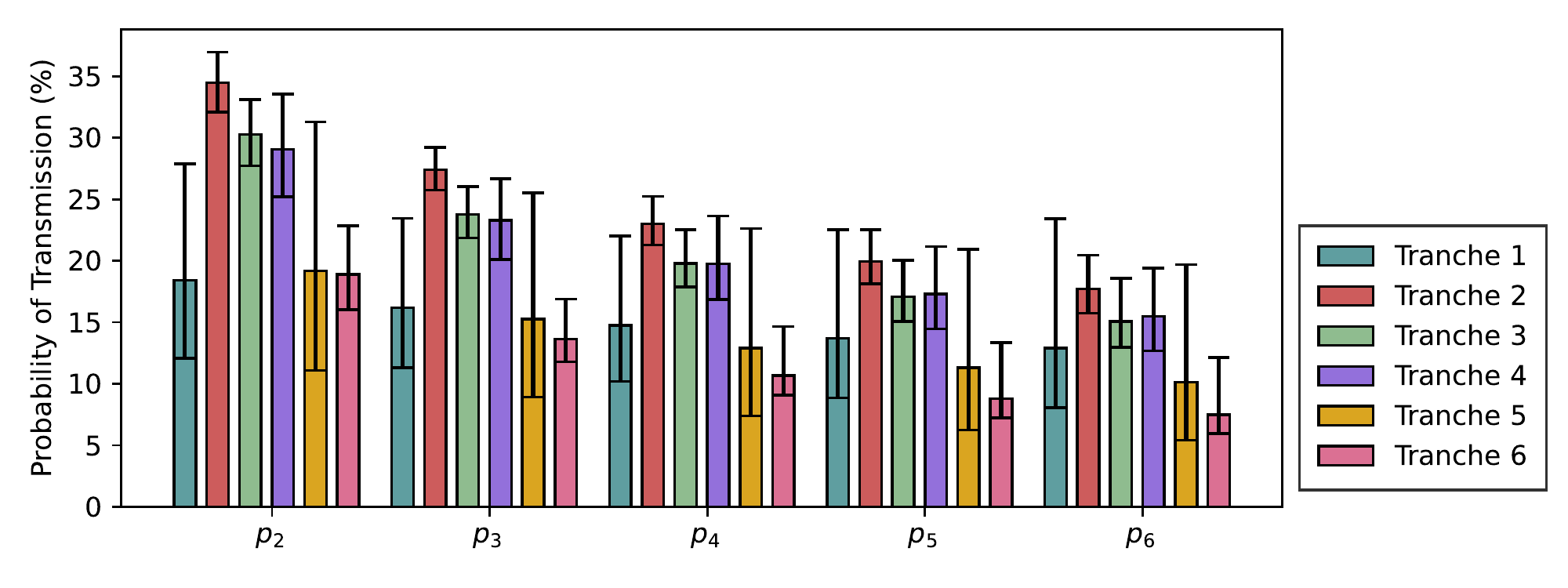}\quad
\end{center}
\caption{Visualisation of the fitted model.  Top: Baseline probability of
infection from outside.  Bottom: Per-pair baseline probabilities of secondary
transmission within the household, not including tertiary transmission
effects.}
\label{fig:model}
\end{figure}

\begin{figure}
\begin{center}
\includegraphics[width=0.8\textwidth]{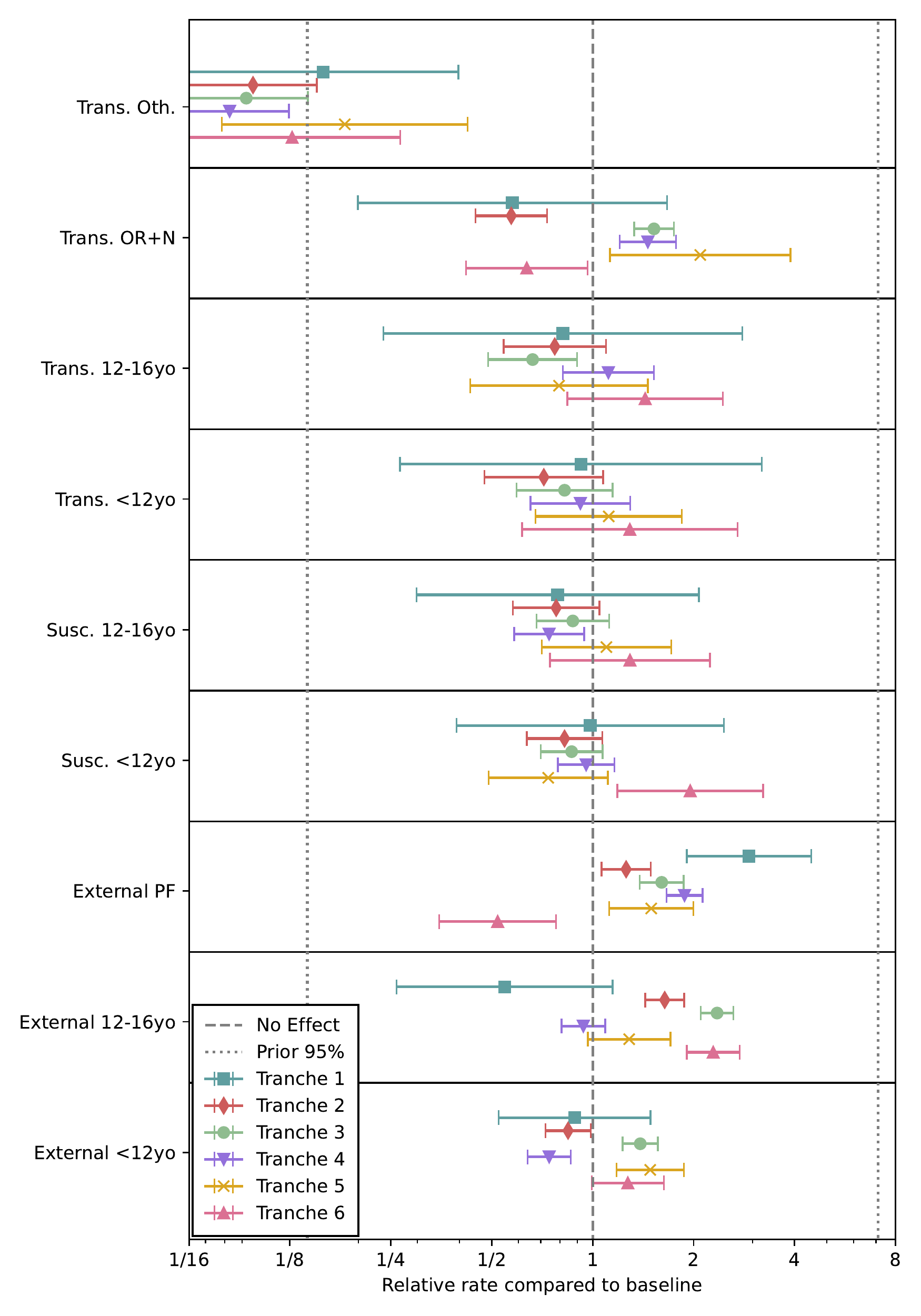}
\end{center}
\caption{Visualisation of the fitted model. Relative effects on transmission,
susceptibility and external exposure compared to baseline of an adult not
working in a patient-facing role with OR+N+S maximal PCR gene positivity pattern if positive.
`Trans.' stands for relative transmissibility, `Susc.' for relative susceptibility,
and `External' for relative external exposure.}
\label{fig:model2}
\end{figure}

The regression analysis has its outputs shown in Table~\ref{tab:fit},
Figure~\ref{fig:model}, and Figure~\ref{fig:model2}. We now present these in
order.

The baseline external probabilities of infection shown in the top plot of
Figure~\ref{fig:model} follow the rough pattern that would be expected from
community prevalence and Tranche duration, with the notable exception of
Tranche 6, when it is likely that vaccination significantly reduced the
infection risk despite high prevalence. In terms of the baseline probabilities
of within-household transmission in the bottom plot of Figure~\ref{fig:model},
these are largely consistent in terms of overlapping credible intervals for
Tranches 2, 3 and 4, with Tranche 6 noticeably lower and with credible
intervals that do not overlap with those for Tranches 2, 3 and 4, likely due to
the impact of vaccination (and despite the emergence of the Delta variant). The
low-prevalence Tranches 1 and 5 have large credible intervals, so are hard to
distinguish statistically from the other Tranches, despite having lower point
estimates. It is worth noting that for periods of low prevalence following
periods of high prevalence, we expect lower viral loads on average as noted by
\citet{Hay:2021}, and this might impact on overall transmissibility estimates.

Turning to Figure~\ref{fig:model2}, we see that `other' patterns of gene
positivity (besides OR+N and OR+N+S) are consistently associated with much
lower transmissibility, as would be expected given target failure is more
likely at lower viral loads\citep{Walker:2020}. We also see lower
transmissibility of OR+N prior to the emergence of the Alpha variant, since
S-gene target failure would have been associated with lower viral loads at that
point as well, but higher transmissibility for this pattern after the emergence
of Alpha but before the emergence of Delta. After the emergence of Delta, the
OR+N pattern is associated with lower transmissibility than OR+N+S, as would be
expected.

In terms of child susceptibility and transmissibility, there is no strong
evidence for an effect. While it is plausible that non-vaccination of children
would lead to increasing their relative susceptibility at later times, this is
consistent with the Tranche 6 results but not strongly evidenced by them. 

For patient-facing staff, external risk of infection has been consistently high
until reduced in Tranche 6, most likely due to the impact of vaccination. For
children, external risk of infection is generally raised compared to baseline
when schools are open, with the exception of primary school aged children
before the emergence of Alpha. Whether this change in association is due to
some causal factor not accounted for here, or is related to the new variants
spreading more efficiently amongst young children than wildtype, requires
further investigation.

\subsection{Limitations and directions for future work}

While we have taken many steps to ensure that the results presented here are as
robust as possible, there are key limitations to the analysis that need to be
borne in mind. The main one of these is failures in ascertainment of positives
and other missingness in the longitudinal design in question. The most likely
consequence of this will be to depress susceptible-infectious transmission
probability estimates. One theoretical approach to deal with this would be
imputation of the transmission tree as suggested by \citet{Demiris:2005}, but
this is likely to be too computationally intensive to be practical in the
current context. Another would be analytical work to include failure of
ascertainment into the likelihood function as in \citet{House:2012}, however it
is unclear how to model ascertainment probabilistically in a tractable manner.
A data-driven approach would be to try to include positives from other sources
such as Test and Trace case data or self-reported episodes of illness. There is
also a harder to quantify potential bias of non-participation in the study,
particularly if this is with respect to some factor that is not measured.

Another important limitation is the possibility that other features, for
example the geographical region that households are in, more detailed
information about viral load and symptoms, or information about the physical
structure of the household, might play an important explanatory role in the
associations observed. Finally, there are possible refinements of the work:
trends in external infection over time could be modelled as a flexible
functional form (e.g.\ a spline as in \citet{Pouwels:2021}); extra features
could be added, and features selected using formal criteria, including relaxing
of the Cauchemez assumption to allow transmission probabilities to depend in a
general manner on household size, and explicit correction to attack rates due
to shrinking and growing epidemics could be made as proposed by
\citeauthor{Ball:2015} \citep{Ball:2015,Shaw:2016}; model parameters -- e.g.\
baseline transmission probabilities -- could be shared across tranches; the
work could be extended to Wales, Scotland and Northern Ireland; more formal
analysis of causal pathways could be performed; and improvements could be made
in implementation data processing, model evaluation through improved linear
algebra, and fitting algorithm. These and other directions should be the
subject of future studies.

\section*{Acknowledgements}

The ONS CIS is funded by the Department of Health and Social Care with in-kind
support from the Welsh Government, the Department of Health on behalf of the
Northern Ireland Government and the Scottish Government.  TH is supported by
the Royal Society (grant number INF/R2/180067). LP is supported by the Wellcome
Trust and the Royal Society (grant number 202562/Z/16/Z). TH, HR and LP are
supported by the UK Research and Innovation COVID-19 rolling scheme (grant
number EP/V027468/1). TH and LR are supported the JUNIPER consortium (grant
number MR/V038613/1) as well as the Alan Turing Institute for Data Science and
Artificial Intelligence.  SB is supported by the Wellcome Trust, the Medical
Research Council and UK Research and Innovation.  KBP and ASW are supported by
the National Institute for Health Research Health Protection Research Unit
(NIHR HPRU) in Healthcare Associated Infections and Antimicrobial Resistance at
the University of Oxford in partnership with Public Health England (PHE)
(NIHR200916).  KBP is also supported by the Huo Family Foundation.  ASW is also
supported by the NIHR Oxford Biomedical Research Centre, by core support from
the Medical Research Council UK to the MRC Clinical Trials Unit
(MC\_UU\_12023/22), and is an NIHR Senior Investigator.  RE is funded by HDR UK
(grant number MR/S003975/1), the MRC (grant number MC\_PC 19065) and the NIHR
(grant number NIHR200908). 

All Authors would like to thank the ONS CIS team. TH and LPwould like to thank
Frank Ball for extremely valuable comments on the work. TH, HR and LP would
like to thank members of the JUNIPER consortium for helpful comments on the
work. The views expressed are those of the authors and not necessarily those of
their employers, funders, the National Health Service, NIHR, Department of
Health, UKHSA, Office for National Statistics, or ONS Data Science Campus.

\begin{landscape}

\begin{table}
\begin{center}
\begin{tabular}{c|c|c|c|c|c|c|c}
Tranche & Start date & End date & Prevalence & Schools & Alpha variant & Delta variant & Vaccination \\
\hline
1 & 26-Apr-20 & 31-Aug-20 & Low & Closed & Not emerged & Not emerged & None \\
2 & 1-Sep-20 & 14-Nov-20 & High & Open & Negligible  & Not emerged & None \\
3 & 15-Nov-20 & 31-Dec-20 & High & Open & Becomes dominant & Not emerged & Negligible \\
4 & 1-Jan-21 & 14-Feb-21 & High & Mainly closed & Dominant & Not emerged & $>10$M 1st, negligible 2nd\\
5 & 15-Feb-21 & 29-Apr-21 & Low & Open & Dominant & Negligible & $>35$M 1st, $>15$M 2nd\\
6 & 30-Apr-21 & 15-Jul-21 & High & Open & Declining & Becomes dominant & $>45$M 1st, $>35$M 2nd\\
\end{tabular}
\end{center}
\caption{Summary of properties of the time periods (tranches) that the data are
split into for analysis.} \label{tab:tranches}
\end{table}

\begin{table}
\begin{center}
\begin{tabular}{c|c|c|c|c|c|c|c}
 & Tranche 1 & Tranche 2  & Tranche 3 & Tranche 4 & Tranche 5 & Tranche 6 & Overall \\
\hline
Number of participants & 89624 & 293570 & 315187 & 329532 & 343821 & 351879 & 408278\\
Number of households & 43300 & 144904 & 157432 & 165238 & 171809 & 178955 & 200876\\
Number of positive individuals & 242 & 5625 & 6078 & 6925 & 1440 & 1890 & 23392\\
Households with $1+$ positive & 206 & 4074 & 4433 & 5123 & 1071 & 1506 & 17180\\
Children $<$12 & 7483 & 23257 & 24045 & 24686 & 25408 & 25050 & 32307\\
Children 12--16 & 4814 & 15503 & 16790 & 18098 & 19012 & 19294 & 22250\\
OR+N+S positives & 124 & 4051 & 2263 & 695 & 33 & 1382 & 9543\\
OR+N positives & 12 & 547 & 2535 & 4353 & 1036 & 244 & 8842\\
Patient-facing participants & 3335 & 9464 & 10046 & 10069 & 11103 & 11437 & 15213
\end{tabular}
\end{center}
\caption{Features of the dataset and different tranches.}
\label{tab:features}
\end{table}

\clearpage

\thispagestyle{empty}

\begin{table}
\begin{center}
\hspace*{-1cm}
\begin{tabular}{c|c|c|c|c|c|c}
 & Tranche 1 & Tranche 2  & Tranche 3 & Tranche 4 & Tranche 5 & Tranche 6 \\
\hline
$1-q$ & 0.237 (0.205,0.274) \% & 1.35 (1.31,1.4) \% & 1.27 (1.22,1.31) \% & 1.56 (1.51,1.61) \% & 0.296 (0.277,0.317) \% & 0.408 (0.384,0.433) \%\\$p_2$ & 18.4 (12.1,27.9) \% & 34.5 (32.1,37.0) \% & 30.2 (27.7,33.1) \% & 29.0 (25.2,33.5) \% & 19.2 (11.1,31.3) \% & 18.9 (16.0,22.9) \%\\$p_3$ & 16.2 (11.3,23.4) \% & 27.4 (25.7,29.2) \% & 23.8 (21.8,26.0) \% & 23.3 (20.1,26.7) \% & 15.3 (8.92,25.5) \% & 13.6 (11.8,16.9) \%\\$p_4$ & 14.8 (10.2,22.0) \% & 23.0 (21.3,25.2) \% & 19.8 (17.9,22.5) \% & 19.7 (16.8,23.6) \% & 12.9 (7.38,22.6) \% & 10.7 (9.09,14.6) \%\\$p_5$ & 13.7 (8.86,22.5) \% & 20.0 (18.1,22.5) \% & 17.1 (15.1,20.0) \% & 17.3 (14.5,21.1) \% & 11.3 (6.25,20.9) \% & 8.79 (7.22,13.3) \%\\$p_6$ & 12.9 (8.06,23.4) \% & 17.7 (15.7,20.5) \% & 15.1 (13.0,18.6) \% & 15.5 (12.7,19.4) \% & 10.1 (5.4,19.7) \% & 7.48 (5.97,12.2) \%\\$\exp(\alpha_{\text{2-11}})$ & 0.883 (0.525,1.49) & 0.845 (0.723,0.987) & 1.39 (1.23,1.56) & 0.742 (0.64,0.86) & 1.48 (1.18,1.87) & 1.27 (0.993,1.63)\\$\exp(\alpha_{\text{12-16}})$ & 0.546 (0.26,1.15) & 1.64 (1.44,1.87) & 2.35 (2.1,2.63) & 0.938 (0.807,1.09) & 1.29 (0.967,1.71) & 2.29 (1.91,2.74)\\$\exp(\alpha_{\text{PF}})$ & 2.93 (1.91,4.49) & 1.26 (1.06,1.49) & 1.61 (1.38,1.87) & 1.88 (1.66,2.13) & 1.5 (1.12,2.0) & 0.521 (0.349,0.778)\\$\exp(\beta_{\text{2-11}})$ & 0.984 (0.393,2.46) & 0.824 (0.636,1.07) & 0.865 (0.7,1.07) & 0.956 (0.787,1.16) & 0.737 (0.49,1.11) & 1.95 (1.18,3.22)\\$\exp(\beta_{\text{12-16}})$ & 0.786 (0.298,2.07) & 0.778 (0.578,1.05) & 0.872 (0.68,1.12) & 0.741 (0.583,0.943) & 1.1 (0.704,1.71) & 1.29 (0.746,2.24)\\$\exp(\gamma_{\text{2-11}})$ & 0.922 (0.266,3.2) & 0.715 (0.476,1.07) & 0.824 (0.593,1.15) & 0.919 (0.652,1.29) & 1.12 (0.676,1.85) & 1.29 (0.615,2.71)\\$\exp(\gamma_{\text{12-16}})$  & 0.815 (0.237,2.8) & 0.771 (0.542,1.1) & 0.662 (0.488,0.899) & 1.11 (0.815,1.52) & 0.794 (0.432,1.46) & 1.43 (0.841,2.45)\\$\exp(\gamma_{\text{OR+N}})$ & 0.576 (0.199,1.67) & 0.572 (0.447,0.731) & 1.52 (1.33,1.75) & 1.46 (1.2,1.77) & 2.09 (1.13,3.89) & 0.636 (0.419,0.965)\\$\exp(\gamma_{\text{CT-oth}})$ & 0.157 (0.062,0.398) & 0.097 (0.0626,0.15) & 0.0926 (0.0607,0.141) & 0.0826 (0.055,0.124) & 0.182 (0.0783,0.424) & 0.127 (0.0604,0.267)
\end{tabular}
\end{center}
\caption{The parameter point estimates and CIs.}
\label{tab:fit}
\end{table}

\end{landscape}

\clearpage

\clearpage

\clearpage

\clearpage

\clearpage

\end{document}